\documentclass[fleqn,usenatbib]{mnras}
\usepackage{general}
\usepackage{mathptmx}
\usepackage{txfonts}
\usepackage[T1]{fontenc}
\usepackage{ae,aecompl}
\newcommand{\rr}{\ensuremath{{\cal R}}}

\title[Nitrogen isotopic ratio of \ce{HC3N} in L1544]{The nitrogen
  isotopic ratio of \ce{HC3N} towards the L1544 prestellar
  core\thanks{Based on observations from the IRAM/ASAI large program}}

\author[P. Hily-Blant]{
P. Hily-Blant,$^{1,2}$\thanks{E-mail: pierre.hily-blant@univ-grenoble-alpes.fr}
A. Faure,$^{2}$
C. Vastel,$^{3}$
V. Magalhaes,$^2$
B. Lefloch,$^{2}$
R. Bachiller$^{4}$
\\
$^1$Institut Universitaire de France\\
$^2$Universit\'e Grenoble Alpes, CNRS, IPAG, F-38000 Grenoble,
France\\
$^3$IRAP, Universit\'e de Toulouse, CNRS, UPS, CNES, Toulouse, France \\
$^4$Observatorio Astron\'omico Nacional (OAN, IGN), Calle Alfonso XII 3, E-28014 Madrid, Spain\\
}

\date{Accepted XXX. Received YYY; in original form ZZZ}

\pubyear{2017}

\hypersetup{draft}
\begin{document}
\def\nratio{\ensuremath{\foun/\fifn}} 
\def\cratio{\ensuremath{\rm ^{12}C/^{13}C}}
\label{firstpage}
\pagerange{\pageref{firstpage}--\pageref{lastpage}}
\maketitle

\begin{abstract}
  The origin of the heavily fractionated reservoir of nitrogen in
  comets remains an issue in the theory of their formation and hence
  of the solar system.  Whether the fractionated reservoir traced by
  comets is inherited from the interstellar cloud or is the product of
  processes taking place in the protostar, or in the protoplanetary
  disk, remains unclear. So far, observations of nitrogen isotopic
  ratios in protostars or prestellar cores have not securely
  identified such a fractionated reservoir owing to the intrinsic
  difficulty of direct isotopic ratios measurements. In this article,
  we report the detection of 5 rotational lines of \ce{HC3N},
  {including the weaker components of the hyperfine multiplets}, and
  two rotational lines of its $^{15}$N isotopologue, towards the L1544
  prestellar core. Based on a MCMC/non-LTE multi-line analysis at the
  hyperfine level, we derive the column densities of \ce{HC3N}
  ($8.0\pm0.4\tdix{13}$\cc) and \ce{HC3^{15}N}
  ($2.0\pm0.4\tdix{11}$\cc) and derive an isotopic ratio of
  400$\pm$20(1$\sigma$).  This value suggests that \ce{HC3N} is
  slightly depleted in $^{15}$N in L1544 with respect to the elemental
  $^{14}$N/$^{15}$N ratio {of $\approx$330} in the present-day local
  interstellar medium. Our study also stresses the need for radiative
  calculations at the hyperfine level. Finally, the comparison of the
  derived ratio with those obtained in CN and HCN in the same core
  seems to favor CN+C$_2$H$_2$ as the dominant formation route to
  HC$_3$N. However, uncertainties in the isotopic ratios preclude
  definitive conclusions.
\end{abstract}


\section{Introduction}

Isotopic ratios are so far the best tool to follow the volatile
content of molecular clouds in their evolution towards the formation
of stars, planets, and planetary systems bodies. Observations of
sublimating ices from comet 67P/C-G with spectrometers on board the
ESA/\textit{ROSETTA} satellite has provided \textit{in situ}
measurements of isotopic ratios of hydrogen, and relative abundances
of \ce{S2} and \ce{O2} molecules which consistently support a presolar
origin for cometary ices \citep{altwegg2015, rubin2015a,
  calmonte2016}. In addition, \textit{Herschel} observations of water
towards protostars suggest that less than 10--20\% of interstellar
ices actually sublimate \citep{vandishoeck2014a}, leaving important
amounts of pristine interstellar material available to build up
planetary systems. These findings motivate the search for an
interstellar heritage within the solar system while providing insights
in the composition of primitive planetary systems in
general. Understanding the link between the chemical composition of
prestellar and protostellar cores and that of comets, which form at
the earliest stages of the planetary formation sequence, is therefore
of paramount importance.

Yet, from an observational perspective, measuring isotopic ratios with
remote observations is a challenging, and intrinsically
time-demanding, task \citep{taniguchi2017c}. Deriving isotopic ratios
generally rests upon several assumptions, some of which being
questionable---such as the co-spatial distribution of
isotopologues---while others are sometimes difficult to
assess---e.g. identical excitation temperatures for the various
isotopologues. From interstellar medium standards, the true accuracy
is usually larger than the typical 5-10\% calibration
uncertainties. In contrast, laboratory or \emph{in situ} measurements
of cosmomaterial samples have accuracy at the percent level
\citep{bonal2010}.

The origin of nitrogen in the solar system is still {an open
  question. More specifically, the main repository of nitrogen in the
  protosolar nebula (PSN) is still unclear, although there is some
  consensus that it may be atomic, N, or molecular, \ce{N2}
  \citep{schwarz2014}. Furthermore, the large variations of the
  isotopic ratio of nitrogen (\nratio), as measured in various
  carriers within different types of solar system objects, remain
  unexplained \citep{aleon2010, furi2015, hilyblant2013a,
    hilyblant2017}.} One striking problem is the \nratio\ isotopic
ratio of nitrogen in comets. Its average value, 144$\pm3$
\citep{jehin2009, bockelee2015, shinnaka2016b, hilyblant2017}, is
three times lower than the bulk ratio of 441$\pm$6 in the protosun as
inferred from solar wind measurements \citep{marty2011}. The reasons
for these different ratios remain elusive, casting doubts on our
understanding of the origin of the composition of comets and more
generally of the origin of nitrogen in the solar system. Several
possibilities (not mutually exclusive) could explain the discrepancy:
\textit{i)} {the tracers of nitrogen observed so far in comets---HCN,
  CN, and \ce{NH2}---are minor reservoirs of cometary nitrogen and
  thus naturally do not reflect the bulk ratio in the PSN},
\textit{ii)} efficient fractionation processes in the protosolar
nebula at the time of comet formation, \textit{iii)} efficient
fractionation processes in the parent interstellar cloud, and
\textit{iv)} exchange processes within cometary ices since their
formation. Recently, it was shown that protoplanetary disks---or
equivalently PSN analogs---carry at least two isotopic reservoirs of
nitrogen, traced respectively by CN and HCN, with HCN probing a
secondary, fractionated, reservoir \citep{hilyblant2017}. Furthermore,
the isotopic reservoirs traced by HCN and CN are found to be in a 1:3
ratio, respectively, reminiscent of the factor of three between the
cometary and bulk isotopic ratios (144:441) in the PSN. It follows
that exchange processes in parent bodies (possibility \textit{iv)}
above) are not necessary. The PSN hypothesis is supported by models of
selective photodissociation of \ce{N2} in protoplanetary disks
\citep{heays2014} which predict a strong enrichment of HCN in \fifn,
but also of CN, in contrast with observations
\citep{hilyblant2017}. At present, clear-cut observational evidences
supporting the PSN or interstellar hypothesis are still lacking.

In this work, we wish to explore the interstellar scenario for the
origin of the heavily fractionated cometary nitrogen. From an
astrophysical perspective, the main issue is observational, in that
the main repositories of nitrogen, N or \ce{N2}, are not directly
observable in cold interstellar clouds. The bulk isotopic ratio is
therefore only measurable indirectly, using the abundances of trace
species such as CN or HNC \citep{adande2012} and chemical models to
infer the bulk. Furthermore, to which extent the isotopic ratio in
these trace molecules is actually representative of the bulk depends
on their detailed formation pathways, especially molecule-specific
fractionation processes which may favor \foun\ or \fifn\ in those
trace species, thus leading to deviations of their isotopic ratios
from that of the bulk. In cold and dense clouds, such processes are
limited to mass fractionation associated to zero-point energy
differences between the two isotopologues \citep{watson1976b,
  terzieva2000, heays2014}. Such processes do apply to deuterium and
hydrogen, where species such as \ce{D2H+} get enriched by orders of
magnitude \citep{vastel2004}. Therefore, to infer the bulk isotopic
ratio from trace species, chemical models including detailed
fractionation reactions must be used \citep{terzieva2000,
  charnley2002, hilyblant2013b, wirstrom2012, roueff2015,
  wirstrom2018}.

To complicate further the inference of isotopic reservoirs in
prestellar cores, the derivation of the abundance of the main
isotopologue, e.g. CN, or HCN, is usually hampered by the large
optical depth of the main isotopologue. Hence, double isotopic ratios
are used, in which the [H\thcn]/[HC\fifn] abundance ratio is measured,
while the [HCN]/[H\thcn] abundance ratio is assumed to be equal to the
carbon elemental isotopic ratio in the local interstellar medium
(ISM), \twc/\thc=68 \citep{milam2005}. This method has been used
intensively, delivering most of the existing nitrogen isotopic ratios
in star forming regions and disks \citep{adande2012, hilyblant2013a,
  hilyblant2013b, wampfler2014, guzman2017, zeng2017,
  colzi2018}. However, detailed model calculations \citep{roueff2015}
have emphasized the pitfalls of the double isotopic ratio method,
especially for HCN, suggesting that [HCN]/[H\thcn] could be larger
than 114, and up to 168, {under typical prestellar core
  conditions}. Recent observations in the L1498 prestellar core also
emphasized the need for direct measurements, although indicating that
the HCN/H\thcn\ ratio is 45$\pm$3\citep{magalhaes2018a}. Isotopic
ratios may also be achieved directly in some instances, e.g. with
doubly substituted molecules such as \ce{NH2D} \citep{gerin2009b}, or
with weak hyperfine lines (CN, HCN) at high signal-to-noise (SNR)
\citep{adande2012, hilyblant2017}. For other species---\ce{N2H+},
\ce{NH3}--- direct ratios have been obtained by means of detailed
radiative transfer models \citep{bizzocchi2013, lis2010, daniel2013,
  daniel2016}, although the derivation remains difficult to assess in
some cases such as HCN \citep{daniel2013}.

In this work, we provide a direct measurement of the nitrogen isotopic
ratio of the \ce{HC3N} cyanopolyyne towards the L1544 prestellar
core. This is the first determination of the \nratio\ ratio of
\ce{HC3N} in a prestellar core. {The L1544 starless core has
  been extensively studied \citep{caselli2012, keto2015, spezzano2017}
  and nitrogen isotopic ratios have been obtained in this source,
  either directly \citep[\ce{N2H+}][]{bizzocchi2013} or through double
  isotopic ratios in HCN or CN \citep{hilyblant2013a,
    hilyblant2013b}. The ratio measured in this work is thus compared
  to that in the chemically related species CN and HCN} to derive some
clues on the formation route(s) of \ce{HC3N} in this source. Our ratio
is also put in perspective with that in \ce{HC3N} and other species
towards different {starless cores located in different
  large-scale environments.} We note, in particular, that the first
\nratio\ ratio of \ce{HC3N} was reported only recently by
\cite{araki2016} in the low-mass star forming region L1527. Even more
recently, this ratio was estimated by \cite{taniguchi2017c} towards
the molecular cloud TMC-1 (cyanopolyyne peak), where HC$_5^{15}$N and
DC$_7$N were also identified \citep{taniguchi2017c, burkhardt2018}.


\begin{figure*}
  \centering
  \includegraphics[width=0.8\hsize]{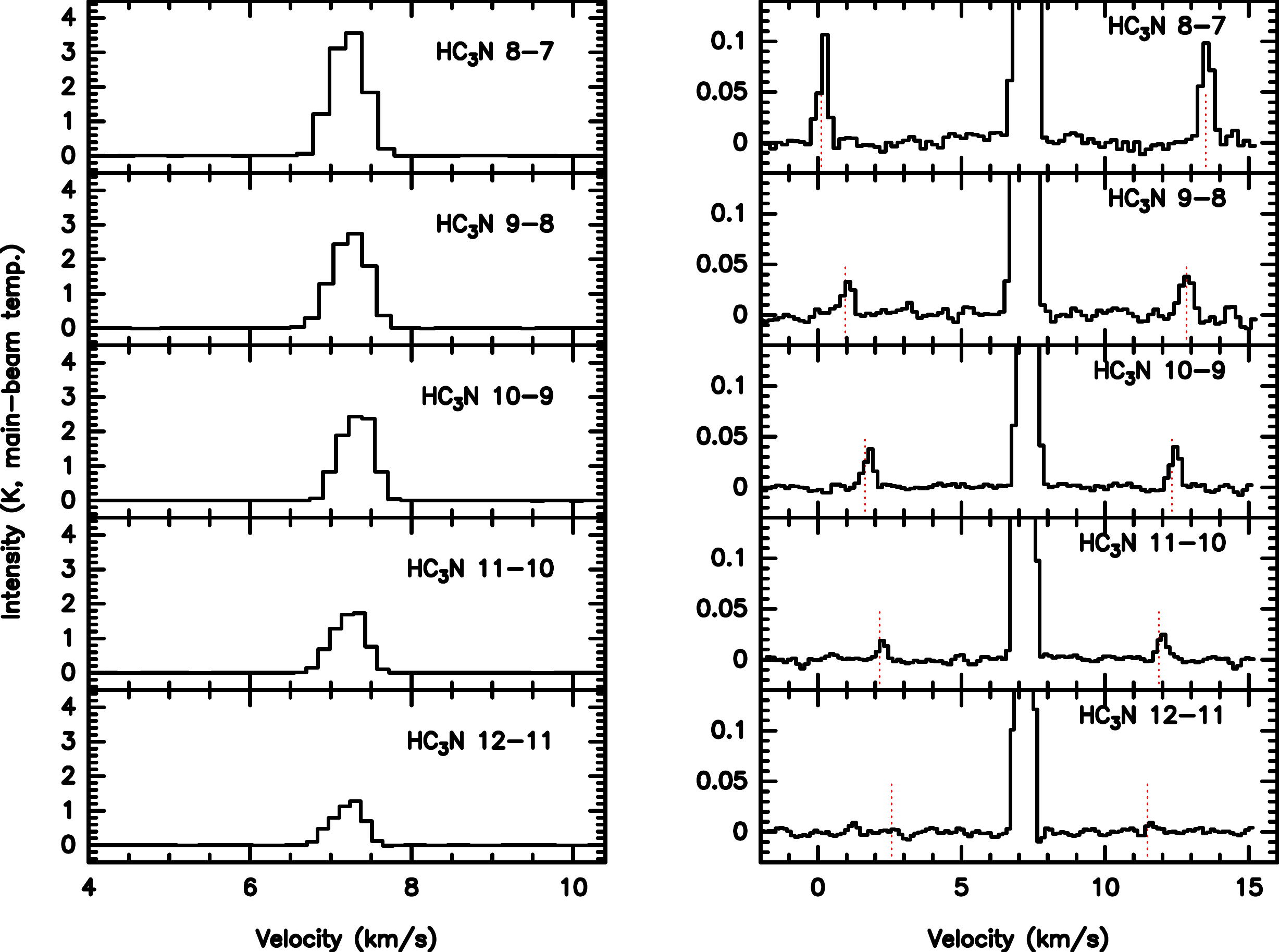}
  \caption{Spectra of \ce{HC3N} towards L1544. The scale is magnified
    in the right panels to show the hyperfine structure.}
  \label{fig:hc3n}
\end{figure*}
\begin{figure}
  \centering
  \includegraphics[width=0.9\hsize]{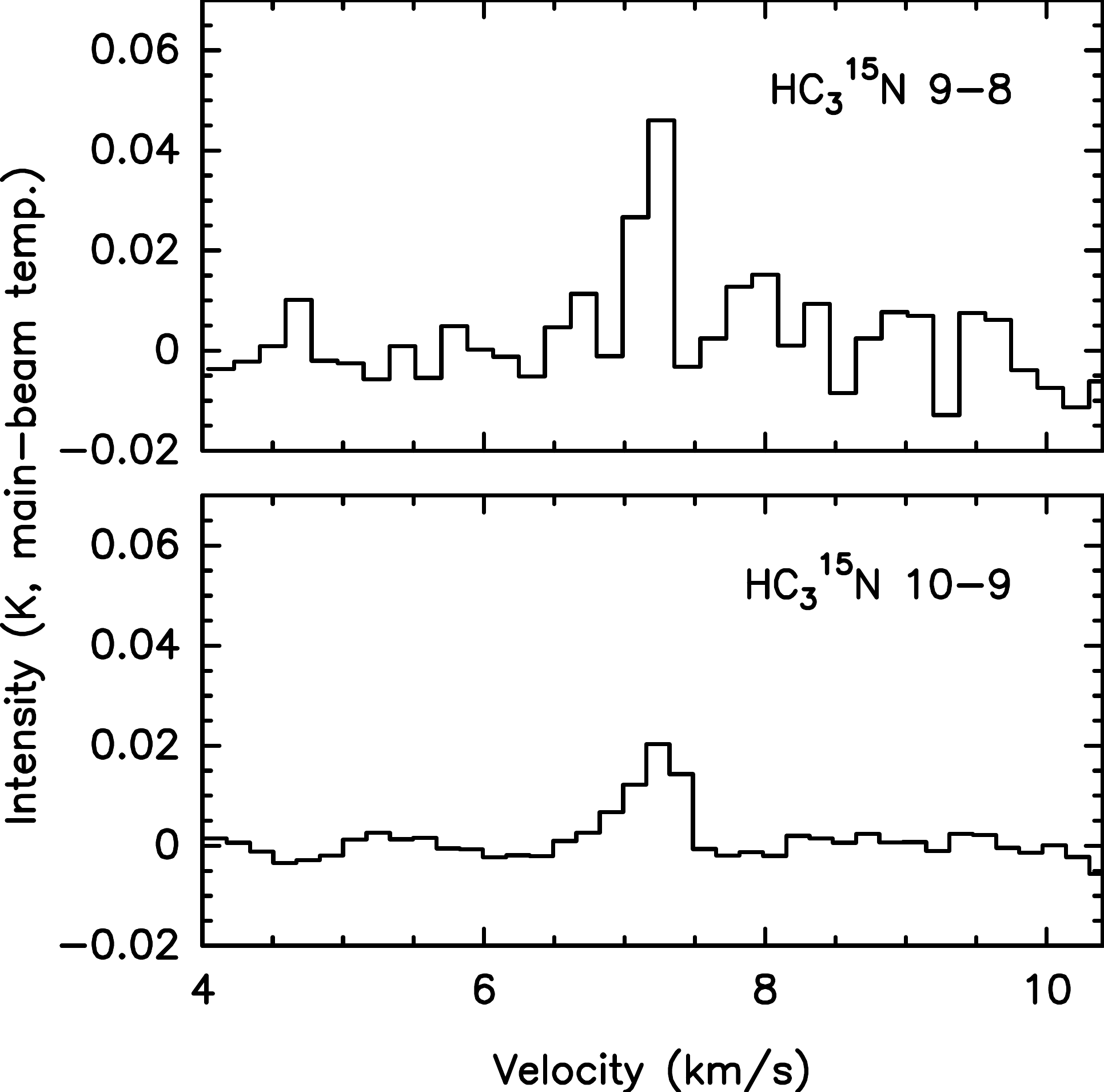}
  \caption{Spectra of the detected \ce{HC3^{15}N} lines.}
  \label{fig:hc315n}
\end{figure}

\section{Observations}
\label{sec:obs}

The observations for all transitions quoted in Table~\ref{tab:obs}
were performed at the IRAM-30m towards the dust peak emission of the
L1544 prestellar core
($\alpha_{2000} = 05^h04^m17.21^s, \delta_{2000} =
25\degr10\arcmin42.8\arcsec$) in the framework of the ASAI Large
Program\footnote{Astrochemical Surveys At IRAM:
  \url{http://www.oan.es/asai}}.  Observations at frequencies lower
than 80~GHz have been performed in December 2015. The EMIR receivers
\citep{carter2012} were used in combination with the Fast Fourier
Transform Spectrometer in its so-called 50~kHz configuration,
providing us with 49~kHz spectral resolution, or typically 0.15 to
0.20 \kms\ at the frequency discussed in this paper. Further details
on the observations can be found in \citet{vastel2014} and
\citet{quenard2017}. {Some, but not all, were observed
  simultaneously. Nevertheless, the lines were observed within short
  period of times thus ensuring consistent amplitude calibration
  throughout the 3mm band. No other lines of \ce{HC3N} or
  \ce{HC3^{15}N} were observed towards L1544 in the frame of the ASAI
  large program.} The forward and beam efficiencies of the
radio-telescope are 0.95 and 0.81 respectively, and all antenna
temperatures were brought into the main-beam temperature scale by
assuming no spatial dilution, namely
$\tmb=\feff/\beff\times\tant$. Contributions from error-beams where
considered negligible for \ce{HC3N}. All the following analysis is
performed in this \tmb\ scale. The spectra were reduced and analyzed
with the \texttt{CLASS} package of the \texttt{GILDAS}
software\footnote{\url{http://www.iram.fr/IRAMFR/GILDAS}}.

\begin{table*}
  \centering
  \caption{\label{tab:obs}Properties of the \ce{HC3N} and
    \ce{HC3^{15}N} lines obtained through Gaussian fitting.}
  \begin{tabular}{llccccccccc}
    \hline
    Molecule & $J'\ra J$ & Rest Frequency$^\P$ & HPBW & $\sigma_T^\S$ & $T^\ddag$
    & $v_0$ & FWHM &
    $W_{\rm main}^\dag$ & $W_{\rm hf}^\dag$ & $W_{\rm tot}^\dag$ \\
    && GHz & \arcsec & mK & K & \kms & \kms & mK \kms & mK \kms & mK \kms\\
    \hline
    \ce{HC3N}
     &  8-7 &    72.78382 & 34 & 5 & 3.78  & 7.22  & 0.51 & 2053 & 88 & 2141\\
     &  9-8 &    81.88147 & 30 & 5 & 2.87  & 7.24  & 0.53 & 1633 & 37 & 1670\\
     & 10-9 &    90.97902 & 27 & 3 & 2.66  & 7.32  & 0.50 & 1404 & 33 & 1437\\
     &11-10 &   100.07639 & 25 & 3 & 1.84  & 7.23  & 0.49 &  956 & 17 &  973\\
     &12-11 &   109.17363 & 23 & 3 & 1.28  & 7.22  & 0.46 &  626 & -- &  626\\
    \hline
    \ce{HC3^{15}N}
     &  9-8 &    79.50005 & 31 & 6 & 0.072 & 7.19 & 0.18 & -- & -- & 14(2)  \\
     & 10-9 &    88.33300 & 28 & 2 & 0.020 & 7.23 & 0.42 & -- & -- & 9.0(7) \\
     &11-10 &    97.16583 & 25 & 3 & --    & --   &  --  & -- & -- & $<$1.4\\
     &12-11 &   105.99852 & 23 & 3 & --    & --   &  --  & -- & -- & $<$1.0\\
     &13-12 &   114.83109 & 21 &14 & --    & --   &  --  & -- & -- & $<$4.3\\
    \hline
  \end{tabular}
  \begin{list}{}{}
  \item[\S] Statistical uncertainty, at the 1$\sigma$ level, on the
    main-beam temperature (in K).
  \item[\P] The rest frequency were taken from the CDMS database.
  \item[\ddag] Peak intensity (main beam temperature).
  \item[\dag] For \ce{HC3N}, three integrated intensity values are
    provided: main component, sum of the two weaker hf lines, and the
    total (main+hf) flux. For \ce{HC3^{15}N}, only the main component
    is detected. The statistical uncertainties (in brackets) are given
    in units of the last digit at the 1$\sigma$ level. Upper limits
    are $1\sigma$.
  \end{list}
\end{table*}
\section{Results}

\subsection{Spectra}

The \ce{HC3N} and \ce{HC3^{15}N} spectra are shown in
Figs.~\ref{fig:hc3n} and \ref{fig:hc315n} respectively. The SNR of the
\ce{HC3N} lines ranges from a few hundreds up to more than a
thousand. Indeed, the hyperfine structure of \ce{HC3N} has been
detected in most cases (see Fig.~\ref{fig:hc3n}), although only the
most widely separated frequencies are resolved leading to essentially
three lines instead of 6. The main line thus consists of 3 overlapping
transitions (see Table~\ref{tab:hc3nhfs}), taking into account that
one hf transition is too low to contribute to the emission. Hyperfine
lines were also reported for the 5-4 transition of \ce{HC3N} towards
L1527 \citep{araki2016}. \rtab{obs} shows that the intensities of the
\ce{HC3^{15}N} lines are typically a hundred times weaker than the
main isotopologue, and only two lines ($J=$9-8 and 10-9) have been
detected.

The observed line properties are summarized in Table~\ref{tab:obs},
including the upper limits on the three undetected \ce{HC3^{15}N}
lines. The integrated intensities of the three sets of hyperfine lines
{of \ce{HC3N}} are given separately in Table~\ref{tab:check}.

Although our chief objective is to directly determine the isotopic
ratio of nitrogen in \ce{HC3N}, the physical conditions of the
emitting gas must also be known to some extent. One important feature
of the present study is the availability of several rotational lines
of each isotopologue, which we analyze simultaneously, providing
strong constraints on the physical conditions and derived column
densities. {Furthermore, the critical density of the \ce{HC3N}
  transitions studied here are between \dix{5} to 5\tdix{5}\ccc, which
  makes them particularly sensitive probes of the density in this, and
  other, starless cores where the density increases from \dix{4} to
  \dix{7}\ccc\ \citep{keto2015}.}

\subsection{Methodology}

Our analysis presented below is based on {non-local thermodynamic
  equilibrium (non-LTE)} radiative transfer calculations. We have used
dedicated collisional rate coefficients for both \ce{HC3N} and
\ce{HC3^{15}N} \citep{faure2016}, extended down to 5~K. The
collisional partner is restricted to ground-state para-H$_2$, which
dominates at {the typical kinetic temperature of 10~K prevailing in
  such core}. For \ce{HC3N}, two sets of rates have been used,
describing the collisions at the rotational and at the hyperfine
levels respectively. For \ce{HC3^{15}N}, only the rotational set is
needed, assuming identical rate coefficients than for
\ce{HC3N}. {Indeed, substitution of $^{14}$N by $^{15}$N reduces the
  rotational constant by only $\sim$3\% while the reduced mass of the
  collisional system is increased by less than 0.1\%. Such changes
  have a negligible impact on the collisional rate coefficients whose
  uncertainty is about 10-20\% \citep{faure2016}.} The non-LTE
radiative transfer calculations were performed using the public code
\texttt{RADEX} \citep{vandertak2007}.

The efficiency of this type of numerical code makes it possible to
perform large grids to explore the four-dimensional parameter space
(density, kinetic temperature, column densities of both species). Such
a frequentist approach allows minima (both local and global) to be
found. However, determining uncertainties based on a simple $\chi^2$
approach usually fails because these parameters are indeed not
independent from each other but are coupled through the complex
interplay of the collisional and radiative excitation processes. In
order to explore, and quantify, the correlations between the
parameters, we adopted the Markov Chain Monte Carlo (MCMC) approach
which enables us to explore the parameter space in a meaningful and
efficient way. In practice, we used the \texttt{emcee} Python
implementation, an affine invariant sampling algorithm, which is
freely available and widely used in the astrophysical
community\footnote{The code is available at
  \url{http://dan.iel.fm/emcee}.}. We also used the \texttt{corner}
Python library to produce plots of the parameter distribution and
correlation matrix.

In this approach, the parameter space is explored using typically 24
Markov chains for typically {1000 to 5000 steps, including a burn-in
  phase of $\sim 300$ steps (see Figs.~\ref{fig:walkershc3nrot} and
  \ref{fig:mcmc15nwalkers}).} In our case, the prior probability was a
uniform distribution for each parameter, with boundaries inferred from
physical and chemical considerations. The kinetic temperature \tkin\
varies between 6 and 15~K, while the log10 of the density, \ce{HC3N}
column density, and isotopic ratio, were taken in the intervals
$[4:7]$, $[11:15]$, and $[2:3]$ respectively. This choice ensures that
the entire ranges of kinetic temperature and density appropriate to
both the envelope and the innermost parts of this core are covered
\citep{keto2015}. The likelihood of each sample was computed as
$\exp(-\chi^2)$, with uncertainties on the integrated intensities
taken to be at least 5\%, significantly larger than the statistical
ones obtained from the Gaussian fitting. In performing the MCMC
parameter space exploration, the number of chains---also called
walkers---and the number of steps of each chain, are critically
important to ensure that the parameter space has been correctly
sampled. It is also important to begin with a physically acceptable
solution, although we have successfully tested that various initial
conditions lead to the same final distributions. By using
multi-threading on a 4 CPU laptop, the typical execution time was 20
min for $10^5$ samples.

\begin{figure*}
  \centering
  \includegraphics[width=.47\hsize]{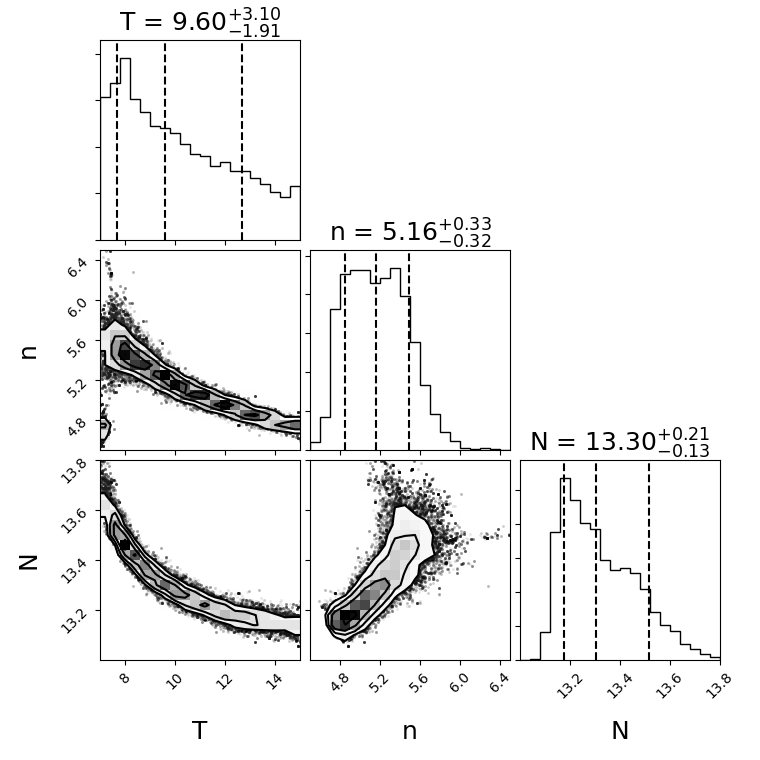}\hfill%
  \includegraphics[width=.47\hsize]{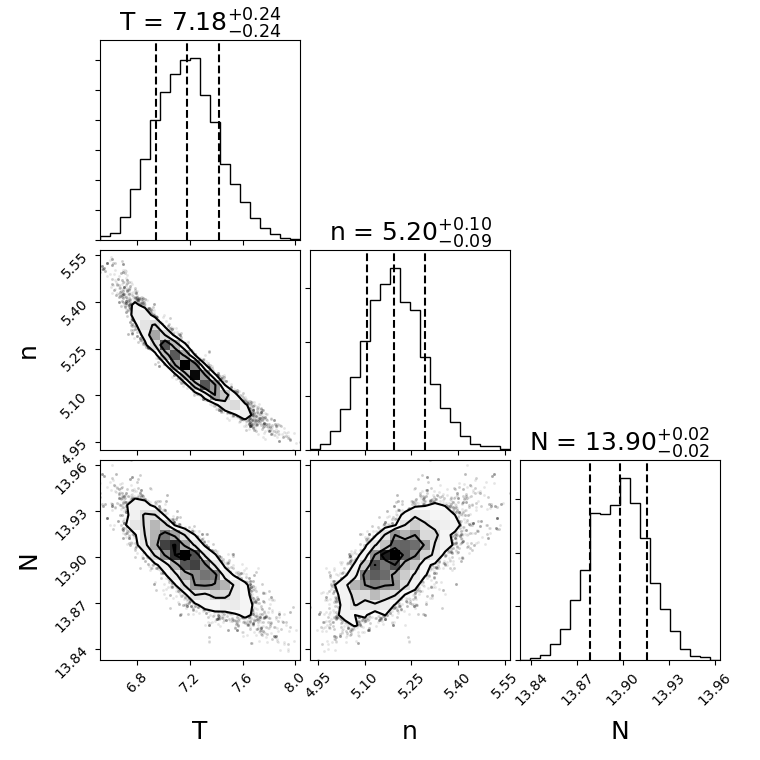}
  \caption{Histograms and cross-histograms from the MCMC parameter
    space exploration, for the kinetic temperature $T$, and for the
    \logd\ of the density ($n$) and column density ($N$) of \ce{HC3N}
    derived from the rotational (left) and hyperfine (right) fluxes
    (see Table~\ref{tab:obs} and Section 3.4). In each histogram
    panel, the median value of the parameter, with the $\pm1\sigma$
    boundaries (16\% and 84\% quantiles), are quoted and indicated by
    vertical dashed-lines (see also Table~\ref{tab:results}).}
  \label{fig:mcmchc3n}
\end{figure*}

\subsection{Assumptions}

The basic assumption is that both isotopologues are co-spatial, not
only in terms of abundance, but also in terms of their excitation,
such that their emissions come from the same regions within the
core. Based on chemical arguments, the two species are expected to be
formed by the same pathways, unless specific fractionation routes
become efficient. In addition, these two molecules have very similar
spectroscopic properties (spontaneous decay and collision rates), and
hence are most likely to emit from the same locations in the cloud.

\subsection{Physical conditions from the \ce{HC3N} lines}

\subsubsection{Rotational analysis}

In a first set of calculations, only the high-SNR \ce{HC3N} lines are
used to determine the physical conditions. {Although the three,
  unresolved, hf components represent more than 99\% of the total
  intensity (see Table~\ref{tab:hc3nhfs}), Gaussian fitting of the
  weaker hf lines show that they carry 2 to 4\% percent of the total
  intensity, thus indicating optical depth effects. The flux
  predictions based on rotational collisional rate coefficients were
  thus compared to the total observed flux ($W_{\rm tot}$ in
  Table~\ref{tab:obs}) of each rotational transition.} The results are
shown in Fig.~\ref{fig:mcmchc3n} and summarized in
Table~\ref{tab:results} (model labeled 'hc3n-rot'). The corresponding
walkers are shown in Appendix (Fig.~\ref{fig:walkershc3nrot}). As is
evident, the five rotational lines do not point towards a single
solution but to an ensemble of parameters. Basically, and not
surprisingly, the physical conditions are slightly degenerated, with
dense gas corresponding to cold temperatures, while warmer gas have
lower density. The density is constrained within an order of
magnitude, from \dix{5} to \dix{6}\ccc, and the kinetic temperature
within 7 to 14~K. In contrast, the column density is relatively well
constrained within 0.4dex,
$N(\ce{HC3N}) = 2.0_{-0.5}^{+1.2}$\tdix{13}\cc. This column density
agrees well with previous analysis \citep{quenard2017}. The presently
determined physical conditions also confirm those reported by these
authors although showing a wider range of solutions.

\subsubsection{Hyperfine analysis}

{A second analysis of the physical conditions of the \ce{HC3N} column
  density was performed in which the flux of each resolved hf line is
  treated separately. The analysis was conducted using the same tools
  but with collision rate coefficients at the hyperfine level. The
  analysis is, however, not as straightforward as in the rotational
  case because, as already mentioned, three hf transitions, each
  contributing $\approx 30\%$ of the total intensity, are not resolved
  (see Table~\ref{tab:hc3nhfs}).} Thus, to compare the RADEX
calculations to the observed flux, the opacity of the three
overlapping hf lines were added and the resulting intensity was
computed, assuming a single excitation temperature, as
\[
  W_{\rm main} = [J_\nu(\texc)-J_\nu(2.73)]\times[1-e^{-\sum_k
    \tau_k}]\times FWHM \times 1.064,
\]
where the sum runs over the 3 overlapping lines and
$J_\nu=h\nu/k/[\exp(h\nu/k\texc)-1]$ is the usual radiation
temperature function. A Gaussian line profile was adopted, an
assumption supported by the opacity of the strongest hf transitions
which remains smaller than 0.5 in all models. We also checked that the
excitation temperatures of the overlapping hf lines within each
rotational transition were effectively equal to within 5\%.

The results of the hyperfine analysis are summarized in
Table~\ref{tab:results} (model labeled 'hc3n-hfs') and shown in
Fig.~\ref{fig:mcmchc3n}. The corresponding walkers are shown in
Appendix (Fig.~\ref{fig:walkershc3nrot}). The correlation between the
kinetic temperature and \hh\ density is tighter than in the rotational
analysis. Kinetic temperatures ranging from 6.5 to 8~K are associated
with \hh\ densities from 2.5 to 1.0\tdix{5}\ccc\, respectively. The
physical conditions are therefore consistent with those from the
rotational analysis, although with a smaller dispersion around the
median values (see Table~\ref{tab:results}). These kinetic
temperatures and densities correspond to the innermost regions of the
model of L1544 of \cite{keto2015}. Nevertheless, the derived column
density of \ce{HC3N} is significantly larger, at the 1$\sigma$ level,
in the hyperfine analysis than from the rotational one, with a value
of $(8.0\pm0.4)$\tdix{13}\cc. The parameters show similar correlations
with each others in both analysis, and in particular the \ce{HC3N}
column density increases with the density, and decreases with the
kinetic temperature.

The excitation temperatures of the best solutions from the rotational
analysis are within 7 to 8~K for all transitions, while somewhat lower
values (5--7~K) are obtained from the hyperfine analysis.

\begin{table*}
  \centering
  \caption{\label{tab:results}Results from the MCMC runs applied to
    the \ce{HC3N} and \ce{HC3^{15}N} fluxes from
    Table~\ref{tab:obs}. Results for different number of steps and/or
    initial conditions are compared.}
  \begin{tabular}{lcc cccc cccc}
    \hline
    RunId & Walkers & Steps
    &\mc{4}{Initial conditions$^\dag$}
    & \mc{4}{Results$^\ddag$} \\
    &&&
    $\tkin$ & $\log \nhh$ & $\log N$ & $\log \rr$&
    $\tkin$ & $\log \nhh$ & $\log N$ & $\log \rr$\\
    &&& K & \ccc & \cc & & K & \ccc & \cc \\
    \hline
    hc3n-rot$^\S$& 24 & 5000
    &8.0&     5.0&    13.1&  -- &9.60(-191,+310)& 5.16(-32,+33)&     13.30(-13,+21)\\
    hc3n-hfs$^\S$ & 24 & 1000 &  8.0&     5.0&    13.1& -- & 7.18(-24,+24)&      5.20(-9,+10)&     13.90(-2,+2)\\
    %
    all-rot$^\sharp$ & 24 & 5000
    & 8.0& 5.6&    13.4& 2.3&   9.44(-183,+339)&  5.24(-36,+35)& 13.30(-14,+22)&  2.32(-5,+ 7)\\
    all-hfs$^\sharp$ & 24 & 5000
    & 8.0& 5.0&    13.1& 2.2&   7.18(-22,+25)&  5.19(-9,+ 9)& 13.89(-2,+ 2)&  2.60(-2,+ 2)\\
    \hline
  \end{tabular}
  \begin{list}{}{}
  \item $^\dag$ $N$ is the total column density of \ce{HC3N} and \rr\ is
    the \ab{HC3N}/\ab{HC3^{15}N} abundance ratio.
  \item $^\ddag$ The median value is given with the 16\% and 84 \%
    quantiles indicated within brackets in units of the last digit.
  \item $^\S$ {Results obtained} from the \ce{HC3N} lines only,
    using either the rotational (hc3n-rot) or hyperfine (hc3n-hfs)
    intensities and the corresponding collision rate coefficients (see
    also Fig.\ref{fig:mcmchc3n}).
  \item $^\sharp$ {Results obtained using both the \ce{HC3N}
      rotational (all-rot) or hf (all-hfs) intensities. See also
      Fig.\ref{fig:mcmchc3n15n}.}
  \end{list}
\end{table*}

\begin{figure*}
  \centering
  \includegraphics[width=0.47\hsize]{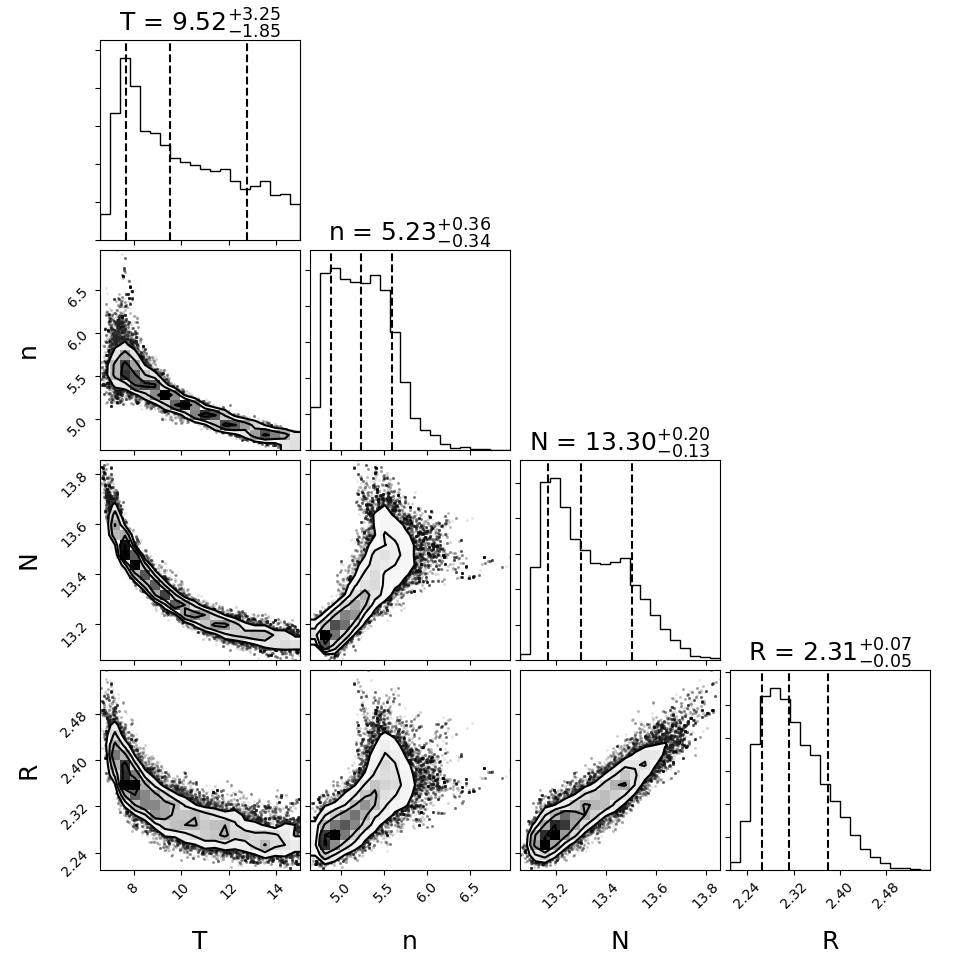}
  \hfill%
  \includegraphics[width=0.47\hsize]{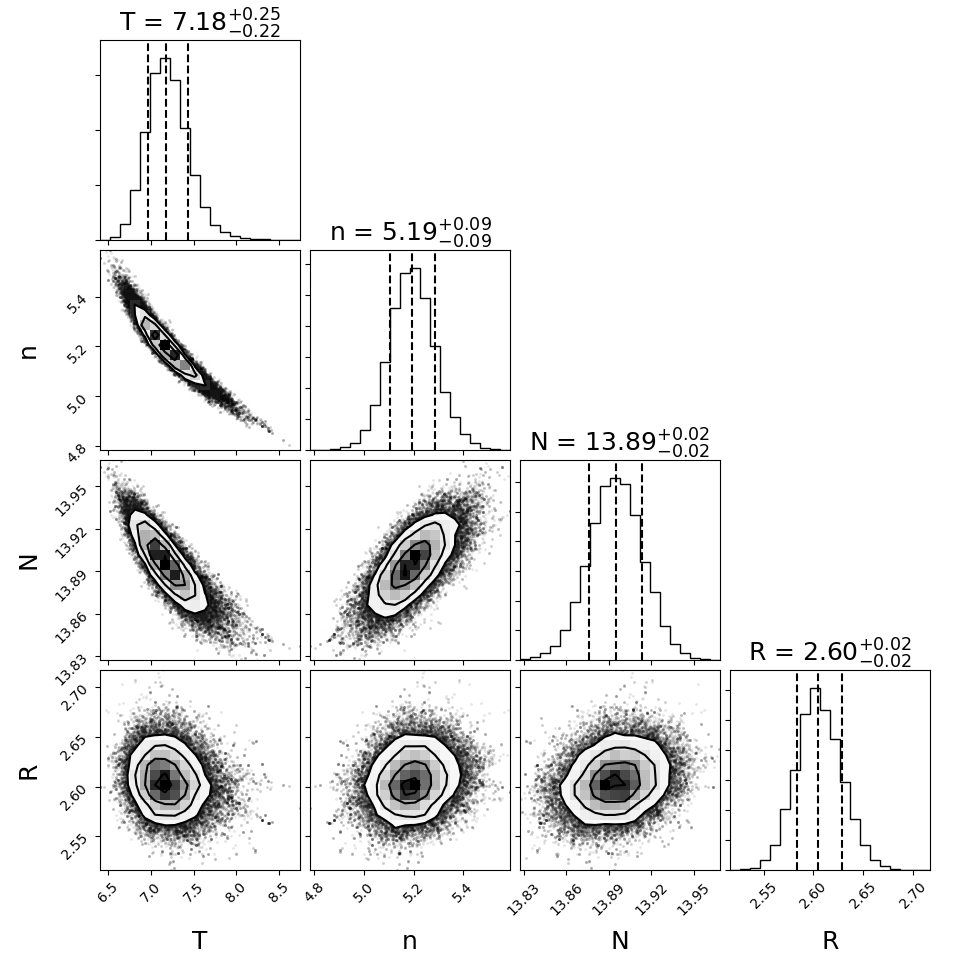}
  \caption{Same as Fig.~\ref{fig:mcmchc3n} for the simultaneous
    analysis of the \ce{HC3N} and \ce{HC3^{15}N} lines, with $R$ the
    \logd\ of the isotopic ratio \rr. The results are summarized in
    Table~\ref{tab:results}. Left: rotational analysis (all-rot
    model). Right: hyperfine analysis (all-hfs model).}
  \label{fig:mcmchc3n15n}
\end{figure*}

\subsection{Nitrogen isotopic ratio in \ce{HC3N}}

To determine the isotopic ratio, the two isotopologues of \ce{HC3N}
were analyzed simultaneously, using either the total rotational fluxes
or the hyperfine fluxes. The upper limits on the \ce{HC3^{15}N} lines
were taken at the 5$\sigma$ level, while the uncertainties for the 5
\ce{HC3N} and 2 \ce{HC3^{15}N} detected lines were, as before, at
least 5\% of the integrated intensity. Several MCMC runs were
performed (not shown), with various numbers of steps and initial
parameters, to check for parameter space exploration and for
convergence towards a consistent set of physical conditions and column
densities. The hyperfine analysis was conducted in the same way as
before, namely by summing the opacities to compute the flux of the
overlapping hyperfine transitions.

The results of the rotational and hyperfine analysis are shown in
Fig.~\ref{fig:mcmchc3n15n} and summarized in Table~\ref{tab:results}
(models all-rot and all-hfs, respectively). The corresponding walkers
are shown in Appendix (Fig.~\ref{fig:mcmc15nwalkers}). Overall, the
parameters show similar correlation patterns as in the previous
analysis. In particular, the physical conditions are consistent with
the above \ce{HC3N} analysis, and the difference between the \ce{HC3N}
column densities from the rotational and hyperfine analysis is
confirmed. {The kinetic temperature are $9.4_{-1.8}^{+3.4}$ and
  7.2$_{-0.2}^{+0.3}$~K from the rotational and hf analysis
  respectively, and are thus only marginally consistent, whereas the
  \ce{H2} densities are fully consistent, with the hf analysis
  resulting in $n(\ce{H2}) = 1.6\pm0.3\tdix{5}\ccc$. The total column
  density of \ce{HC3N} from the hf analysis ($7.8\pm0.4\tdix{13}$\cc)
  is again significantly larger than from the rotational analysis
  ($2.0_{-0.5}^{+1.3}\tdix{13}\cc$). These differences translate into
  the significantly distinct isotopic ratios}: the rotational and
hyperfine analysis lead to $\rr=\ab{HC3N}/\ab{HC3^{15}N}$=216$\pm$30
and 400$\pm$20 respectively. The corresponding \ce{HC3^{15}N} column
densities differ by a factor two, being
$(9.5_{-3.4}^{+9.1})\tdix{10}$\cc\ and
$(19.5_{-1.7}^{+1.9})$\tdix{10}\cc\ respectively.

To compare the two sets of solutions, we computed the hyperfine fluxes
associated with the most probable parameters obtained from each
analysis (Table~\ref{tab:results}). These predictions are then
compared to the observed fluxes by considering their algebraic distance
from the observed flux measured in units of the $1\sigma$ rms. The
results are shown in Fig.~\ref{fig:results}. The flux of the strongest
{hyperfine lines of each rotational line} of \ce{HC3N} lines
are well reproduced, within $\pm3\sigma$, by both analysis. The fluxes
and upper limits of \ce{HC3^{15}N} are also well matched. However, the
flux of the two weaker hyperfine lines of the 8-7 transition are
under-predicted by more than 10$\sigma$ by the rotational
analysis. Similar discrepancies are also found for the 10-9 and 11-10
transitions. In comparison, the hyperfine analysis solution is able to
reproduce all fluxes to within $\pm5\sigma$. The underestimate of the
weak, optically thin, hyperfine lines by the rotational analysis
directly translates into an underestimate of the column
density---despite a larger kinetic temperature---of the main
isotopologue, hence, of the isotopic ratio. As is evident, the
hyperfine solution is to be preferred.

Another difference between the two analysis is the correlation between
$N(\ce{HC3N})$ and \rr\ obtained in the rotational fitting but not in
the hyperfine analysis (see Fig.~\ref{fig:mcmchc3n15n}). Instead, the
isotopic ratio presents no clear correlation with any other
parameter. This difference suggests that handling separately the
hyperfine fluxes adds significant constraints which force the system
towards higher \ce{HC3N} column densities and removes the low
values. Finally, we note that the excitation temperatures are similar
to those obtained in the previous analysis of \ce{HC3N} only. In
particular, the excitation temperatures are close for both
isotopologues.

In summary, from our non-LTE hyperfine analysis, the resulting
\ab{HC3N}/\ab{HC3^{15}N} abundance ratio in L1544 is:
\begin{equation*}
  \rr=400\pm20.
  \label{eq:rr}
\end{equation*}
{The quoted uncertainty is calculated from the 16\% and 84\%
  quantiles (i.e. $\pm1\sigma$) of the resulting distribution as
  obtained from the MCMC sampling. This uncertainty indeed corresponds
  to the adopted conservative 5\% calibration uncertainty.}

\section{Discussion}

\subsection{Nitrogen fractionation in \ce{HC3N}} 

\begin{figure}
  \centering
  \includegraphics[width=\hsize]{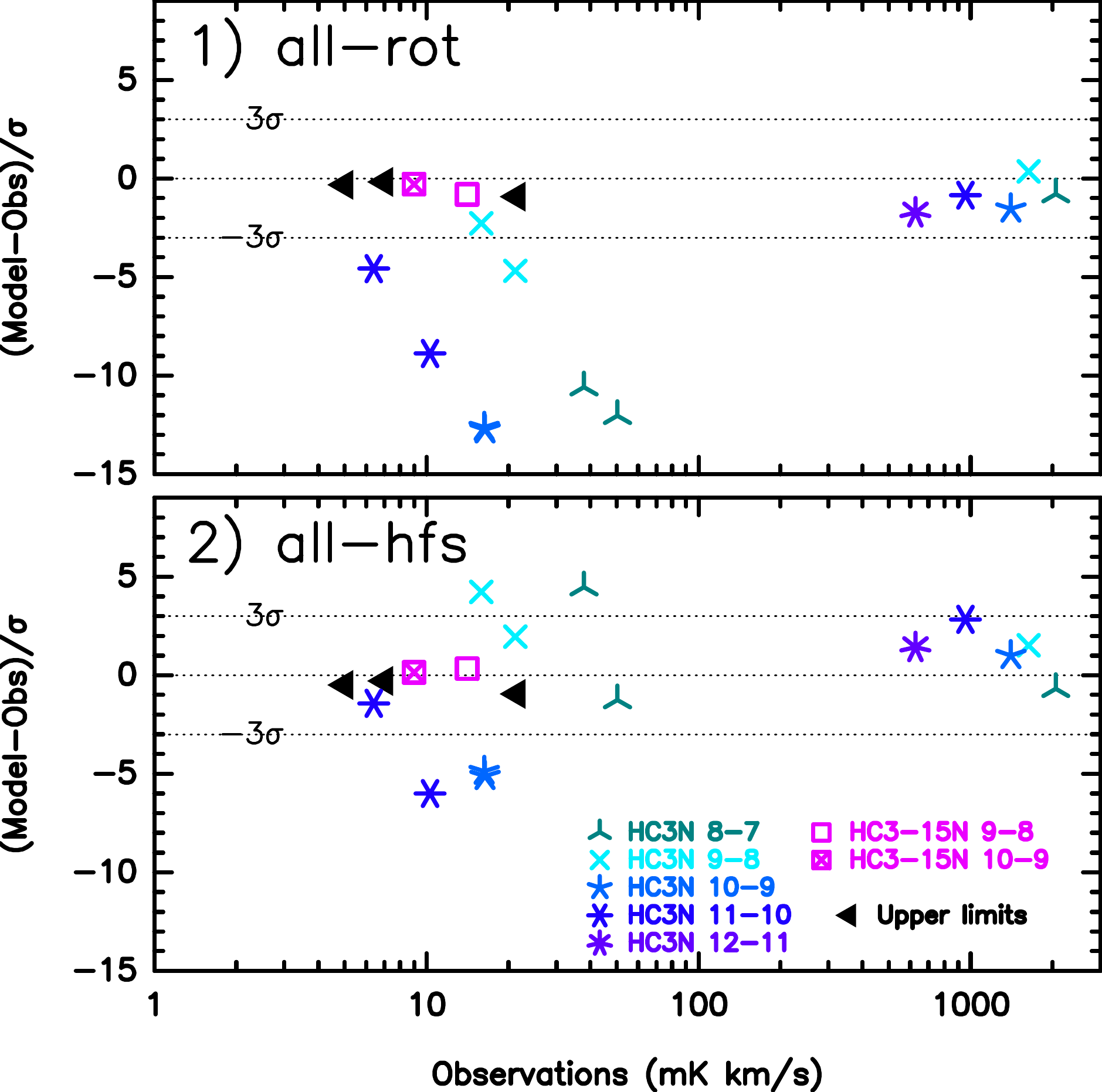}
  \caption{Comparison of the observed and predicted integrated
    hyperfine intensities (in units of the rms, $\sigma$) for the best
    solutions obtained from the rotational (panel 1, all-rot) and
    hyperfine (panel 2, all-hfs) analysis, as reported in
    Tables~\ref{tab:results} and \ref{tab:check}. In each panel, the
    fluxes of the hyperfine components are given for all \ce{HC3N}
    lines except the 12-11 transition for which only the main hf line
    was detected (see Table~\ref{tab:obs}). The filled triangles
    indicate the 5$\sigma$ upper limits on the \ce{HC3^{15}N}
    rotational transitions.}
  \label{fig:results}
\end{figure}

When compared to the elemental ratio of 323$\pm30$ in the solar
neighborhood \citep{hilyblant2017}, the present direct determination
of the \ab{HC3N}/\ab{HC3^{15}N} ratio\footnote{{It should be stressed
    that the isotopic ratios mentioned in what follows are beam
    averaged measurements.}} in L1544 shows that \ce{HC3N} is, at the
$1\sigma$ level, slightly depleted in \fifn. Recently, the nitrogen
isotopic ratio of \ce{HC3N} was measured towards the cyanopolyynes
peak (CP) in TMC-1 (see \rtab{compil}) and a value of $257\pm54$ was
obtained \citep{taniguchi2017c}, not consistent with our present
determination.  Because both cores are located within the solar
neighborhood, the elemental isotopic ratio of nitrogen should be
identical, and different isotopic ratios in L1544 and TMC-1 CP thus
indicate different dominant chemical formation pathways in both
cores. We note, however, that the TMC1-CP measurement relies on the
double isotopic ratio using \thc\ isotopologues of \ce{HC3N}, and a
single rotation line ($J=4\to 3$) of \ce{HC3^{15}N} is employed.

{In the same core, these authors obtained a direct measurement of the
  \ce{HC5N}:\ce{HC5^{15}N} abundance ratio which was found to be
  323$\pm$80, hence indicating no fractionation of nitrogen in this
  molecule. An indirect measurement by the same authors lead to
  344$\pm$80. Both measurements thus agree with our value in L1544 for
  \ce{HC3N} to within 1$\sigma$ uncertainties, which could be
  interpreted as a signature of common chemical pathways for these two
  cyanopolyynes, in contrast with the conclusion of these
  authors. Direct measurement of \nratio\ in \ce{HC3N} in TMC1-P,
  using the same kind of approach as the present one, would help
  clarifying the issue.}

The \nratio\ isotopic ratio in \ce{HC3N} can also be compared to
ratios derived in other nitrogen bearing molecules, namely CN, HCN and
\ce{N2H+}, in the L1544 prestellar core (\rtab{compil}). For CN and
HCN, the ratios were derived indirectly using the double isotopic
method, leading to $510\pm70$ for CN and to within 140 to 360 for HCN
\citep{hilyblant2013a, hilyblant2013b}. The present result,
$\rr=400\pm20$, is thus marginally consistent with these values and
therefore suggests that CN, HCN and \ce{HC3N} could share a common
nitrogen reservoir, as discussed in more detail below. In \ce{N2H+}
towards L1544, the ratio is $1000\pm200$, which is also the largest
ratio found in any species towards prestellar cores
\citep{hilyblant2017}. The ratios in \ce{N2H+} and \ce{HC3N} are
therefore significantly different, at the 3$\sigma$ level, indicating
that they either sample different pools of nitrogen atoms, or that
so-far unknown fractionation reactions are unevenly partitioning
$^{14}$N and $^{15}$N among these two species.

\begin{table}
  \caption{Compilation of nitrogen isotopic ratios in cyanopolyynes
    and the directly related species CN and HCN.}
\label{tab:compil}
  \begin{tabular}{llclr}
    \toprule
    Source   & Species   &  ${\cal R}$ & Method$^\S$ & Reference \\
    \midrule
    TMC1(CP) & \ce{HC3N} & 270$\pm$57 & Direct & (1--3)\\
             & \ce{HC3N} & 257$\pm$54 & Indirect & (1--2)\\
             & \ce{HC5N} & 323$\pm$80 & Direct & (1)\\
             & \ce{HC5N} & 344$\pm$80 & Indirect & (1) \\
    L1527    & \ce{HC3N} & 338$\pm$12 & Indirect & (4) \\
    L1544    & \ce{HC3N} & 400$\pm$20 & Direct & This work \\
             & HCN       & 140--350   & Indirect & (5) \\
             & CN        & 500$\pm$75 & Indirect & (6) \\
    L1498    & HCN       & 338$\pm$28 & Direct   & (7) \\
             & CN        & 500$\pm$75 & Indirect & (6) \\
    TW Hya   & CN        & 323$\pm$30 & Direct\\
    \bottomrule
  \end{tabular}
  \begin{list}{}{}
  \item $\S$ Direct methods measure the [X\fifn]/[X\foun] abundance
    ratio; indirect methods use double isotopic ratios.
  \item References: (1) \cite{taniguchi2017c} (2) \cite{kaifu2004} (3)
    \cite{takano1998} (4) \cite{araki2016} (5) \cite{hilyblant2013a}
    (6) \cite{hilyblant2013b} (7) \cite{magalhaes2018a}
  \end{list}
\end{table}

\subsection{Uncertainties}

Obtaining reliable and as small as possible uncertainties on column
density ratios is certainly one of the main challenges in remote
isotopic ratio measurements such as the one presented here.

The statistical uncertainties on the observed \ce{HC3N} integrated
intensities are extremely small (Table~\ref{tab:obs}). Yet, they
hardly reflect the total uncertainty which must take into account the
influence of the weather, pointing, focus, and chopper wheel
calibration method. At the IRAM-30m telescope, the overall calibration
uncertainty is generally taken to be 5\% to 10\%. Given the fact that
all lines in our survey were observed in a consistent way and over a
relatively short periods of time, we adopted a flux calibration
uncertainty of 5\%.

One way to mitigate amplitude calibration uncertainties is to perform
simultaneous observations of different lines, which naturally cancel
out the multiplicative fluctuations, such as receiver
gains. Nonetheless, differential effects across the receiver bandpass
remain. Furthermore, it is usually assumed that the main-beam
temperature is the appropriate temperature scale, but the true
brightness temperature should indeed be derived by convolving the
source structure by the complete telescope beam (not only the
main-beam). Yet, for species such as \ce{HC3N}, one does not expect
extended, strong, emission, which could contribute significantly
through error-beam pick-up.

From an observational perspective, it therefore seems extremely
difficult to go beyond the calibration-limited accuracy on isotopic
ratios. From this regard, our conservative approach differs from that
of other authors who propagate the statistical uncertainties
\citep{araki2016} and claim to reach a 1-2\% uncertainty level,
despite the use of lines obtained with different telescopes (GBT and
NRO).

Furthermore, in addition to the flux calibration, line analysis
requires assumptions to be made which carry additional sources of
uncertainties---which amount essentially to the assumptions in the
radiative transfer---that are usually very difficult to quantify,
unless several models can be compared. For instance, the LTE
assumption\footnote{For consistency with the common usage, we use LTE
  to refer to the assumption of a single (excitation) temperature to
  describe the level population of a (set of) molecule(s), which is
  indeed a weaker assumption than that of local thermodynamical
  equilibrium.}  is commonly used to compute column densities from
single lines \citep{hilyblant2013a, araki2016,
  taniguchi2017c}. Although the uncertainty associated to the assumed
value of the excitation temperature can be quantified---and is
actually generally small when dealing with column density ratios---the
uncertainty associated to the LTE assumption itself is much more
difficult to estimate. Strictly speaking, the uncertainty on the
derived excitation temperature should be propagated into the column
densities and column density ratios. Yet, in their analysis,
\citeauthor{araki2016} apply the single excitation temperature
assumption to the main and \thc\ isotopologues although the excitation
temperatures of the \thc\ isotopomers differ by up to 20\% (see their
Table~2). Their very small uncertainties on the column density ratios
thus neglects the uncertainty associated to their single excitation
temperature assumption.

In the present analysis, we have relaxed this assumption by computing
non-LTE level populations, at the rotational and hyperfine
levels. However, our calculations assume uniform physical conditions
although it is clear that assumptions on the source geometry represent
an additional source of uncertainties. Indeed, we note that, although
the hyperfine fitting provides a much better overall agreement (see
\rfig{results}), some weak hyperfine components are reproduced only at
the 5$\sigma$ level, which may indicate the limits of our 0D
model. The impact of this assumption can only be tested through
multi-dimensional, non-LTE, calculations combined with spectral line
mapping observations \citep{daniel2013, magalhaes2018a}.

\subsection{Formation of \ce{HC3N}}
\def\tcch{\ce{^{13}CCH}}
\def\ctch{\ce{C^{13}CH}}
\def\hca{\ce{H^{13}CCCN}}
\def\hcb{\ce{HC^{13}CCN}}
\def\hcc{\ce{HCC^{13}CN}}

The chemistry of \ce{HC3N} in dense prestellar cores is still a matter
of debate although it has been studied in great detail especially
using carbon isotopic ratios. In particular, $^{13}$C isotopic
anomalies were observed for \ce{HC3N} in the Taurus Molecular Cloud-1
(TMC-1) starless core towards the so-called Cyanopolyyne Peak (CP)
\citep{takano1998}. The column densities of \hca, \hcb, and \hcc\
follow the ratios {1.0:1.0:1.4}, respectively, which was interpreted
as a result of the production pathway of \ce{HC3N} rather than isotope
exchange reactions. The latter are unlikely because the zero-point
energy differences between the three carbon isotopologues would not
predict [\hca]/[\hcb] close to 1. Instead, \ce{HC3N} is thought to
form primarily from a parent molecule having two equivalent carbon
atoms in order to explain [\hca]=[\hcb]. Currently, three potential
formation routes are considered, through CN, HCN, and HNC \citep[for a
review, see e.g.][]{taniguchi2016}.

\subsubsection{The \ce{HCN} route}

In this route, the two equivalent carbons are provided by \ce{C2H2+}:
\begin{equation}
  \ce{C2H2+ + HCN -> HC3NH+ + H}.
  \label{hcn}
\end{equation}
This reaction is known to be rapid at room temperature with a rate
coefficient $k_1 =3.78\times 10^{-11}$\cccs\ \citep{iraqi1990}. This
reaction is followed by the dissociative recombination (DR) between
\ce{HC3NH+} and electrons. The branching ratios of this DR process
have not been measured for each product channel \citep{vigren2012} but
quantum calculations indicate that all isomers of \ce{HC3N}, namely
\ce{HNC3}, HCCNC and HCNCC can be produced at low temperature. In
particular, if reaction~(\ref{hcn}) is a major pathway to produce
\ce{HC3N}, then \ce{HC3N} and \ce{HNC3} are expected to have similar
abundances since both products result from a direct hydrogen
elimination of the parent ion \ce{HC3NH+} \citep{osamura1999}. The
observation of \ce{HC3N} and its isomers HCCNC and \ce{HNC3}, however,
indicates that the three species have very different abundances in the
cold conditions of TMC-1 \citep{ohishi1998} as well as in L1544
\citep{vastel2018}. Assuming that \ce{HC3N} and \ce{HNC3} have a
similar reactivity, this suggests that reaction~(\ref{hcn}) is not a
major pathway to \ce{HC3N} \citep[see the discussion in][]{vastel2018}.

\subsubsection{The CN route}

Here, the two equivalent carbons come from \ce{C2H2}:
\begin{equation}
  \ce{C2H2 + CN -> HC3N + H}
  \label{cn}
\end{equation}
This reaction is also rapid at room temperature with a rate
coefficient $k_2= 2.5\times 10^{-10}$~\cccs\
\citep{sims1993}. Reaction~(\ref{cn}) has even been found to
accelerate down to 25~K \citep{sims1993}. The current consensus is
that \ce{HC3N} is mainly produced through the neutral-neutral
reaction~(\ref{cn}). In addition the larger abundance of
\ce{HCC^{13}CN} relative to \ce{H^{13}CCCN} and \ce{HC^{13}CCN} can be
easily explained by different $^{12}$C/$^{13}$C isotopic ratios in
\ce{C2H2} and CN, since the triple -CN bond is conserved in
reaction~(\ref{cn}).
This scenario has been recently supported by observations of \hca,
\hcb, \hcc, and \ce{HC3N} towards the L1527 protostar, which were
found to follow the ratios 1.00:1.01:1.35:86.4, respectively
\citep{araki2016}. This corresponds to a
$^{12}$C/$^{13}$C isotopic ratio of 64 for
[HCC$^{13}$CN]/[\ce{HC3N}], in good agreement with the local ISM
elemental
$^{12}$C/$^{13}$C ratio of 68 \citep{milam2005}. This in turn seems to
indicate that CN is not fractionated in carbon.
Furthermore, the significantly higher \cratio\ ratios in \hca\ and
\hcb\ of $\approx$86 is consistent with
$^{13}$C depletion in \ce{C2H2}, as observed in other carbon-chain
molecules \citep[see][and references therein]{roueff2015, araki2016}.

In the L1527 protostar, \cite{araki2016} also determined a
[HCC$^{13}$CN]/[HC$_3^{15}$N] abundance ratio of
5.26$\pm$0.19, resulting in a
[\ce{HC3N}]/[HC$_3^{15}$N] ratio of 338$\pm
12$. This value is again close to the elemental \nratio\ ratio in the
local ISM \citep{adande2012, hilyblant2017}. To our knowledge the
\cratio\ and \nratio\ ratios in CN in L1527 are unknown so that a firm
conclusion regarding the link between CN and \ce{HC3N} cannot be
established in this source.

\subsubsection{The HNC route}

The third route, with three non-equivalent carbon atoms, is:
\begin{equation}
  \ce{HNC + C2H -> HC3N + H}.
  \label{hnc}
\end{equation}
The rate coefficient is unknown but the {\it ab initio} computations
of \cite{fukuzawa1997} have shown that this reaction is exothermic
with no energy barrier. A rate coefficient of
$k_3= 1.75\times 10^{-10}$~\cccs\ was suggested by
\cite{hebrard2012}. In reaction~(\ref{hnc}), the H atom is removed
from HNC, and we thus expect [\hca]/[\hcb]=[\ctch]/[\tcch], at odds
with the observations towards TMC-1 and L1527 which find
[\ctch]/[\tcch] abundance ratios of $\sim 1.6$ \citep{sakai2010}. The
origin of the departure of [\ctch]/[\tcch] from unity is likely
associated to the neutral-neutral fractionation reaction \ce{^{13}CCH
  + H -> C^{13}CH + H + $8.1$K}. Indeed, at a kinetic temperature of
15~K, the equilibrium abundance ratio due to this reaction is
$\exp(8.1/15)=1.7$, which is in harmony with the observed ratios. At
the lower temperature of TMC-1 (CP), where the steady-state ratio
would be 2.2, chemical model calculations show that the
1.6$\pm0.4(3\sigma)$ ratio can be obtained at early times, when carbon
is still mostly neutral \citep{furuya2011}. As a consequence of the
different [\hca]/[\hcb] and [\ctch]/[\tcch] abundance ratios, the HNC
route was discarded by \cite{taniguchi2016}.

However, recently, the [\hca]/[\hcb] abundance ratio was measured
towards two cold prestellar cores, L1521B and L134N, with values of
0.98$\pm0.14$ and 1.5$\pm0.2$ (1$\sigma$ uncertainties), respectively,
suggesting different formation pathways for \ce{HC3N} in these two
sources \citep{taniguchi2017b}. In particular, the ratio in L134N
could indicate that \ce{HC3N} is primarily formed from \ce{HNC + C2H},
while \ce{CN + C2H2} would be the main route in L1521B, as in
TMC-1(CP) and L1527 \citep{takano1998, araki2016}. The competition
between the CN and HNC routes was interpreted by \cite{taniguchi2017b}
as reflecting different [CN]/[HNC] abundance ratios in the two
prestellar cores L1521B and L134N. Observations of the [\ctch]/[\tcch]
in both sources should bring valuable information.

\subsubsection{Consequences of the nitrogen isotopic ratio of \ce{HC3N}}

The present determination of the \ab{HC3N}/\ab{HC3^{15}N} ratio in
L1544 brings a new piece of the \ce{HC3N} puzzle. The [CN]/[C\fifn]
abundance ratio in this source is 510$\pm$70, as obtained using a
non-LTE hyperfine analysis and assuming a [CN]/[$^{13}$CN] ratio of 68
\citep{hilyblant2013b}. The \nratio\ ratio in \ce{HC3N} is thus
consistent with the [CN]/[C\fifn] ratio in L1544 to within
$1.5\sigma$.


On the other hand, [HCN]/[HC\fifn]=$257\pm30$ was measured at the
central position of L1544, again using a non-LTE hyperfine analysis
and assuming a value of 68 for the [HCN]/[H$^{13}$CN] ratio
\citep{hilyblant2013a}. The \nratio\ ratio in \ce{HC3N} is thus not
consistent with that of [HCN]/[HC$^{15}$N] in L1544, which would
exclude the HCN route, in agreement with the HC$_3$N isomer abundances
discussed above. We note, however, that the uncertainties introduced
by the double isotopic method (for both CN and HCN) are certainly
larger than the above quoted error bars. Finally, to our knowledge, no
measurement of \nratio\ in HNC has been reported so far in L1544. It
is therefore not possible to conclude with certainty about the main
production pathway of \ce{HC3N} in L1544: each of the three above
routes (from HCN, CN and HNC) might indeed contribute to the derived
\nratio\ ratio of 400. The HCN route is, however, very likely a minor
pathway.

\subsection{Fractionation of nitrogen in prestellar cores}

{Our \ab{HC3N}/\ab{HC3^{15}N} abundance ratio of 400 suggests
  that \ce{HC3N} is only slightly depleted in $^{15}$N with respect to
  the elemental ratio of $\approx$330 in the solar
  neighborhood.}

{Models of nitrogen chemical fractionation have been developed and
  applied to dense gas, from the typical \dix{4}\ccc\ of prestellar
  clouds where mild effects are predicted \citep{terzieva2000}, to the
  much higher density, \dix{7}\ccc, of protostars \citep{charnley2002,
    wirstrom2018} where high degrees of fractionation are promoted by
  the heavy depletion of carbon monoxide. More recently, the set of
  fractionation reactions was revised \citep{roueff2015}, and the
  dominant fractionation routes were disqualified on energetic
  grounds. Accordingly, these models predict that the steady-state
  isotopic ratios of all trace species should reflect the elemental
  ratio. Regardless of the detailed formation pathway of \ce{HC3N},
  our new direct measurement thus seems to support these model
  predictions. However, the current understanding of nitrogen chemical
  fractionation is still not well established. Indeed, chemical
  calculations using the fractionation network of \cite{terzieva2000}
  in combination with the chemical network of \cite{hilyblant2010n}
  predict significant variations of the isotopic ratios for most of
  the observable nitrogenated species, especially HCN
  \citep{hilyblant2013b}. Moreover, the isotopic ratio measured
  directly in \ce{N2H+} is $\approx 1000$ in the L1544 core,
  suggesting strong depletion in \fifn\ \citep{bizzocchi2013} which no
  model is able to reproduce. Finally, we note that the recent models
  of \citeauthor{roueff2015} also predict that carbon fractionation
  would lead to HCN/H\thcn\ ratio up to $\sim$140, in sharp contrast
  with the value of 45$\pm$3 obtained recently in the L1498 prestellar
  core \citep{magalhaes2018a}. Therefore, putting these model
  predictions on a firmer ground, requires further observations not
  only to accurately measure isotopic ratios but also to test the
  chemical fractionation models.}

\section{Concluding remarks}

Molecular isotopic ratios are invaluable tools for studying the origin
of the solar system and the possible link between the primordial
matter and interstellar chemistry. As discussed above, they also offer
powerful diagnostics to reveal the underlying chemistry. We have
derived the first $^{14}$N/$^{15}$N ratio of \ce{HC3N} in a prestellar
core, L1544. This ratio, $\rr=400\pm20$, is consistent with that
derived in the same source for CN ($510\pm70$) {but
  significantly larger than the ratio derived indirectly for HCN
  ($257\pm30$)}. As such, the present observations seem to favor the
\ce{CN + C2H2} route as the major pathway to \ce{HC3N}. A similar
conclusion was also reached by \cite{vastel2018} based on the relative
abundances of the HC$_3$N isomers.

{From a chemical modelling perspective, two important issues
  must be addressed. One is the discrepancy with the observations
  regarding the \cratio\ in HCN, which may be related to incorrect
  fractionation routes, or to the chemistry of nitriles itself. The
  other is the \nratio\ ratio in \ce{N2H+}. At this point, it must be
  emphasized that observational constraints to fractionation models
  are provided in terms of the comparison of the measured isotopic
  ratios in a given species against the elemental ratio. The elemental
  ratio is known to vary with time and with the galactocentric
  distance \citep{adande2012, colzi2018, romano2017}. The value of the
  elemental \nratio\ ratio in the local ISM, at $\approx 8$ kpc, to
  which measurements in the local ISM, such as the present
  \ce{HC3N}/\ce{HC3^{15}N} ratio should be compared, is expected to be
  lower than its value in the PSN, 4.6~Gyr ago, based on galactic
  chemical evolution model predictions \citep{romano2017}. A new value
  of this ratio, $\approx$330, was indeed proposed for the present-day
  solar neighborhood \citep{hilyblant2017}, based on the compilation
  of direct measurements obtained in the dense, local, ISM. This value
  also agrees very well with the model predictions
  \citep{romano2017}. In addition, new measurements have been
  published recently that are in very good agreement with this
  elemental ratio \citep{taniguchi2017c, kahane2018,
    magalhaes2018a}. Because chemical mass fractionation decreases
  with temperature, observations of warm gas provide more direct
  probes of the bulk ratio in the present-day local ISM.}

{From an observational perspective, it is still necessary to
  disentangle between the different precursors of HC$_3$N. In
  particular, the measurement of the \cratio\ ratios in the three
  $^{13}$C isotopomers of HC$_3$N would provide a crucial test of the
  CN versus HNC routes. The determination of the \nratio\ ratio in
  \ce{HC3NH+} would also be decisive to definitively exclude the HCN
  route. We note that the main isotopologue, \ce{HC3NH+}, has been
  observed recently in L1544 \citep{quenard2017}.}

{More generally, further observations and radiative transfer
  calculations are required to put robust constraints on the \nratio\
  isotopic ratio in various species, including hydrides and nitriles,
  in L1544 and similar prestellar cores. In particular, future works
  should revisit the \nratio\ in CN and HCN in the L1544 (and other)
  cores, as recently done in L1498 \citep{magalhaes2018a}. In this
  context, we emphasize that future isotopic ratios should be
  determined with a methodology similar to the one presented in this
  work, i.e. a non-LTE analysis combined with a robust statistical
  approach in order to control, and minimize, the
  uncertainties. Moreover, the present work demonstrates the large,
  non-statistical, uncertainties resulting from using a rotational
  analysis. This stresses the need for collision rate coefficients at
  the hyperfine level. Ideally, hyperfine overlap must be taken into
  account, as was recently demonstrated for HCN in L1498
  \cite{magalhaes2018a}.}


\section*{Acknowledgments}

We thank the anonymous referee for a careful reading and general
comments which helped to improve the manuscript. We wish to thank
Martin Legrand and Luc Lefort who participated into the multi-line
analysis and the implementation of the MCMC/Radex code during their
internship in our group in May-July 2017. PHB acknowledges the
\emph{Institut Universitaire de France} for financial support. AF
acknowledges the financial support from the CNRS Programme National
PCMI (Physique et Chimie du Milieu Interstellaire).

\bibliographystyle{mnras}
\input{hc3n_publisher.bbl}

\begin{thebibliography}{}
\makeatletter
\relax
\def\mn@urlcharsother{\let\do\@makeother \do\$\do\&\do\#\do\^\do\_\do\%\do\~}
\def\mn@doi{\begingroup\mn@urlcharsother \@ifnextchar [ {\mn@doi@}
  {\mn@doi@[]}}
\def\mn@doi@[#1]#2{\def\@tempa{#1}\ifx\@tempa\@empty \href
  {http://dx.doi.org/#2} {doi:#2}\else \href {http://dx.doi.org/#2} {#1}\fi
  \endgroup}
\def\mn@eprint#1#2{\mn@eprint@#1:#2::\@nil}
\def\mn@eprint@arXiv#1{\href {http://arxiv.org/abs/#1} {{\tt arXiv:#1}}}
\def\mn@eprint@dblp#1{\href {http://dblp.uni-trier.de/rec/bibtex/#1.xml}
  {dblp:#1}}
\def\mn@eprint@#1:#2:#3:#4\@nil{\def\@tempa {#1}\def\@tempb {#2}\def\@tempc
  {#3}\ifx \@tempc \@empty \let \@tempc \@tempb \let \@tempb \@tempa \fi \ifx
  \@tempb \@empty \def\@tempb {arXiv}\fi \@ifundefined
  {mn@eprint@\@tempb}{\@tempb:\@tempc}{\expandafter \expandafter \csname
  mn@eprint@\@tempb\endcsname \expandafter{\@tempc}}}

\bibitem[\protect\citeauthoryear{{Adande} \& {Ziurys}}{{Adande} \&
  {Ziurys}}{2012}]{adande2012}
{Adande} G.~R.,  {Ziurys} L.~M.,  2012, \mn@doi [\apj]
  {10.1088/0004-637X/744/2/194}, \href
  {http://adsabs.harvard.edu/abs/2012ApJ...744..194A} {744-758, 194}

\bibitem[\protect\citeauthoryear{{Al{\'e}on}}{{Al{\'e}on}}{2010}]{aleon2010}
{Al{\'e}on} J.,  2010, \mn@doi [\apj] {10.1088/0004-637X/722/2/1342}, \href
  {http://adsabs.harvard.edu/abs/2010ApJ...722.1342A} {722, 1342}

\bibitem[\protect\citeauthoryear{{Altwegg} et~al.,}{{Altwegg}
  et~al.}{2015}]{altwegg2015}
{Altwegg} K.,  et~al., 2015, \mn@doi [Science] {10.1126/science.1261952}, 347

\bibitem[\protect\citeauthoryear{{Araki}, {Takano}, {Sakai}, {Yamamoto},
  {Oyama}, {Kuze}  \& {Tsukiyama}}{{Araki} et~al.}{2016}]{araki2016}
{Araki} M.,  {Takano} S.,  {Sakai} N.,  {Yamamoto} S.,  {Oyama} T.,  {Kuze} N.,
    {Tsukiyama} K.,  2016, \mn@doi [\apj] {10.3847/1538-4357/833/2/291}, \href
  {http://adsabs.harvard.edu/abs/2016ApJ...833..291A} {833, 291}

\bibitem[\protect\citeauthoryear{{Bizzocchi}, {Caselli}, {Leonardo}  \&
  {Dore}}{{Bizzocchi} et~al.}{2013}]{bizzocchi2013}
{Bizzocchi} L.,  {Caselli} P.,  {Leonardo} E.,   {Dore} L.,  2013, \mn@doi
  [\aap] {10.1051/0004-6361/201321276}, \href
  {http://adsabs.harvard.edu/abs/2013A%26A...555A.109B} {555, A109}

\bibitem[\protect\citeauthoryear{{Bockel{\'e}e-Morvan}
  et~al.,}{{Bockel{\'e}e-Morvan} et~al.}{2015}]{bockelee2015}
{Bockel{\'e}e-Morvan} D.,  et~al., 2015, \mn@doi [\ssr]
  {10.1007/s11214-015-0156-9}, \href
  {http://adsabs.harvard.edu/abs/2015SSRv..197...47B} {197, 47}

\bibitem[\protect\citeauthoryear{{Bonal}, {Huss}, {Krot}, {Nagashima}, {Ishii}
  \& {Bradley}}{{Bonal} et~al.}{2010}]{bonal2010}
{Bonal} L.,  {Huss} G.~R.,  {Krot} A.~N.,  {Nagashima} K.,  {Ishii} H.~A.,
  {Bradley} J.~P.,  2010, \mn@doi [\gca] {10.1016/j.gca.2010.08.017}, \href
  {http://adsabs.harvard.edu/abs/2010GeCoA..74.6590B} {74, 6590}

\bibitem[\protect\citeauthoryear{{Burkhardt}, {Herbst}, {Kalenskii},
  {McCarthy}, {Remijan}  \& {McGuire}}{{Burkhardt}
  et~al.}{2018}]{burkhardt2018}
{Burkhardt} A.~M.,  {Herbst} E.,  {Kalenskii} S.~V.,  {McCarthy} M.~C.,
  {Remijan} A.~J.,   {McGuire} B.~A.,  2018, \mn@doi [\mnras]
  {10.1093/mnras/stx2972}, \href
  {http://adsabs.harvard.edu/abs/2018MNRAS.474.5068B} {474, 5068}

\bibitem[\protect\citeauthoryear{Calmonte et~al.,}{Calmonte
  et~al.}{2016}]{calmonte2016}
Calmonte U.,  et~al., 2016, \mn@doi [\mnras] {10.1093/mnras/stw2601}, 462, S253

\bibitem[\protect\citeauthoryear{{Carter} et~al.,}{{Carter}
  et~al.}{2012}]{carter2012}
{Carter} M.,  et~al., 2012, \mn@doi [\aap] {10.1051/0004-6361/201118452}, \href
  {http://adsabs.harvard.edu/abs/2012A%26A...538A..89C} {538, A89}

\bibitem[\protect\citeauthoryear{{Caselli} et~al.,}{{Caselli}
  et~al.}{2012}]{caselli2012}
{Caselli} P.,  et~al., 2012, \mn@doi [\apjl] {10.1088/2041-8205/759/2/L37},
  \href {http://adsabs.harvard.edu/abs/2012ApJ...759L..37C} {759, L37}

\bibitem[\protect\citeauthoryear{{Charnley} \& {Rodgers}}{{Charnley} \&
  {Rodgers}}{2002}]{charnley2002}
{Charnley} S.~B.,  {Rodgers} S.~D.,  2002, \mn@doi [\apjl] {10.1086/340484},
  \href {http://adsabs.harvard.edu/abs/2002ApJ...569L.133C} {569, L133}

\bibitem[\protect\citeauthoryear{{Colzi}, {Fontani}, {Caselli}, {Ceccarelli},
  {Hily-Blant}  \& {Bizzocchi}}{{Colzi} et~al.}{2018}]{colzi2018}
{Colzi} L.,  {Fontani} F.,  {Caselli} P.,  {Ceccarelli} C.,  {Hily-Blant} P.,
  {Bizzocchi} L.,  2018, \mn@doi [\aap] {10.1051/0004-6361/201730576}, \href
  {http://adsabs.harvard.edu/abs/2018A%26A...609A.129C} {609, A129}

\bibitem[\protect\citeauthoryear{{Daniel} et~al.,}{{Daniel}
  et~al.}{2013}]{daniel2013}
{Daniel} F.,  et~al., 2013, \mn@doi [\aap] {10.1051/0004-6361/201321939}, \href
  {http://adsabs.harvard.edu/abs/2013A%26A...560A...3D} {560, A3}

\bibitem[\protect\citeauthoryear{{Daniel} et~al.,}{{Daniel}
  et~al.}{2016}]{daniel2016}
{Daniel} F.,  et~al., 2016, \mn@doi [\aap] {10.1051/0004-6361/201628192}, \href
  {http://cdsads.u-strasbg.fr/abs/2016A%26A...592A..45D} {592, A45}

\bibitem[\protect\citeauthoryear{{Faure}, {Lique}  \& {Wiesenfeld}}{{Faure}
  et~al.}{2016}]{faure2016}
{Faure} A.,  {Lique} F.,   {Wiesenfeld} L.,  2016, \mn@doi [\mnras]
  {10.1093/mnras/stw1156}, \href
  {http://adsabs.harvard.edu/abs/2016MNRAS.460.2103F} {460, 2103}

\bibitem[\protect\citeauthoryear{{Fukuzawa} \& {Osamura}}{{Fukuzawa} \&
  {Osamura}}{1997}]{fukuzawa1997}
{Fukuzawa} K.,  {Osamura} Y.,  1997, \mn@doi [\apj] {10.1086/304782}, \href
  {http://adsabs.harvard.edu/abs/1997ApJ...489..113F} {489, 113}

\bibitem[\protect\citeauthoryear{{F{\"u}ri} \& {Marty}}{{F{\"u}ri} \&
  {Marty}}{2015}]{furi2015}
{F{\"u}ri} E.,  {Marty} B.,  2015, \mn@doi [Nature Geosc.] {10.1038/ngeo2451},
  \href {http://adsabs.harvard.edu/abs/2015NatGe...8..515F} {8, 515}

\bibitem[\protect\citeauthoryear{{Furuya}, {Aikawa}, {Sakai}  \&
  {Yamamoto}}{{Furuya} et~al.}{2011}]{furuya2011}
{Furuya} K.,  {Aikawa} Y.,  {Sakai} N.,   {Yamamoto} S.,  2011, \mn@doi [\apj]
  {10.1088/0004-637X/731/1/38}, \href
  {http://adsabs.harvard.edu/abs/2011ApJ...731...38F} {731, 38}

\bibitem[\protect\citeauthoryear{{Gerin}, {Marcelino}, {Biver}, {Roueff},
  {Coudert}, {Elkeurti}, {Lis}  \& {Bockel{\'e}e-Morvan}}{{Gerin}
  et~al.}{2009}]{gerin2009b}
{Gerin} M.,  {Marcelino} N.,  {Biver} N.,  {Roueff} E.,  {Coudert} L.~H.,
  {Elkeurti} M.,  {Lis} D.~C.,   {Bockel{\'e}e-Morvan} D.,  2009, \mn@doi
  [\aap] {10.1051/0004-6361/200911759}, \href
  {http://adsabs.harvard.edu/abs/2009A%26A...498L...9G} {498, L9}

\bibitem[\protect\citeauthoryear{{Guzm{\'a}n}, {{\"O}berg}, {Huang}, {Loomis}
  \& {Qi}}{{Guzm{\'a}n} et~al.}{2017}]{guzman2017}
{Guzm{\'a}n} V.~V.,  {{\"O}berg} K.~I.,  {Huang} J.,  {Loomis} R.,   {Qi} C.,
  2017, \mn@doi [\apj] {10.3847/1538-4357/836/1/30}, \href
  {http://adsabs.harvard.edu/abs/2017ApJ...836...30G} {836, 30}

\bibitem[\protect\citeauthoryear{{Heays}, {Visser}, {Gredel}, {Ubachs},
  {Lewis}, {Gibson}  \& {van Dishoeck}}{{Heays} et~al.}{2014}]{heays2014}
{Heays} A.~N.,  {Visser} R.,  {Gredel} R.,  {Ubachs} W.,  {Lewis} B.~R.,
  {Gibson} S.~T.,   {van Dishoeck} E.~F.,  2014, \mn@doi [\aap]
  {10.1051/0004-6361/201322832}, \href
  {http://adsabs.harvard.edu/abs/2014A%26A...562A..61H} {562, A61}

\bibitem[\protect\citeauthoryear{{H{\'e}brard}, {Dobrijevic}, {Loison},
  {Bergeat}  \& {Hickson}}{{H{\'e}brard} et~al.}{2012}]{hebrard2012}
{H{\'e}brard} E.,  {Dobrijevic} M.,  {Loison} J.~C.,  {Bergeat} A.,   {Hickson}
  K.~M.,  2012, \mn@doi [\aap] {10.1051/0004-6361/201218837}, \href
  {http://adsabs.harvard.edu/abs/2012A%26A...541A..21H} {541, A21}

\bibitem[\protect\citeauthoryear{{Hily-Blant}, {Walmsley}, {Pineau des
  For{\^e}ts}  \& {Flower}}{{Hily-Blant} et~al.}{2010}]{hilyblant2010n}
{Hily-Blant} P.,  {Walmsley} M.,  {Pineau des For{\^e}ts} G.,   {Flower} D.,
  2010, \mn@doi [A\&A] {10.1051/0004-6361/200913200}, \href
  {http://adsabs.harvard.edu/abs/2010A%26A...513A..41H} {513, A41}

\bibitem[\protect\citeauthoryear{{Hily-Blant}, {Bonal}, {Faure}  \&
  {Quirico}}{{Hily-Blant} et~al.}{2013a}]{hilyblant2013a}
{Hily-Blant} P.,  {Bonal} L.,  {Faure} A.,   {Quirico} E.,  2013a, \mn@doi
  [\icarus] {10.1016/j.icarus.2012.12.015}, \href
  {http://adsabs.harvard.edu/abs/2013Icar..223..582H} {223, 582}

\bibitem[\protect\citeauthoryear{{Hily-Blant}, {Pineau des For{\^e}ts},
  {Faure}, {Le Gal}  \& {Padovani}}{{Hily-Blant}
  et~al.}{2013b}]{hilyblant2013b}
{Hily-Blant} P.,  {Pineau des For{\^e}ts} G.,  {Faure} A.,  {Le Gal} R.,
  {Padovani} M.,  2013b, \mn@doi [\aap] {10.1051/0004-6361/201321364}, \href
  {http://adsabs.harvard.edu/abs/2013A%26A...557A..65H} {557, A65}

\bibitem[\protect\citeauthoryear{{Hily-Blant}, {Magalhaes}, {Kastner}, {Faure},
  {Forveille}  \& {Qi}}{{Hily-Blant} et~al.}{2017}]{hilyblant2017}
{Hily-Blant} P.,  {Magalhaes} V.,  {Kastner} J.,  {Faure} A.,  {Forveille} T.,
   {Qi} C.,  2017, \mn@doi [\aap] {10.1051/0004-6361/201730524}, \href
  {http://adsabs.harvard.edu/abs/2017A%26A...603L...6H} {603, L6}

\bibitem[\protect\citeauthoryear{{Iraqi}, {Petrank}, {Peres}  \&
  {Lifshitz}}{{Iraqi} et~al.}{1990}]{iraqi1990}
{Iraqi} M.,  {Petrank} A.,  {Peres} M.,   {Lifshitz} C.,  1990, \mn@doi
  [International Journal of Mass Spectrometry and Ion Processes]
  {10.1016/0168-1176(90)85102-8}, \href
  {http://adsabs.harvard.edu/abs/1990IJMSI.100..679I} {100, 679}

\bibitem[\protect\citeauthoryear{{Jehin}, {Manfroid}, {Hutsem{\'e}kers},
  {Arpigny}  \& {Zucconi}}{{Jehin} et~al.}{2009}]{jehin2009}
{Jehin} E.,  {Manfroid} J.,  {Hutsem{\'e}kers} D.,  {Arpigny} C.,   {Zucconi}
  J.-M.,  2009, \mn@doi [Earth Moon and Planets] {10.1007/s11038-009-9322-y},
  \href {http://adsabs.harvard.edu/abs/2009EM%26P..105..167J} {105, 167}

\bibitem[\protect\citeauthoryear{{Kahane}, {Jaber Al-Edhari}, {Ceccarelli},
  {L{\'o}pez-Sepulcre}, {Fontani}  \& {Kama}}{{Kahane}
  et~al.}{2018}]{kahane2018}
{Kahane} C.,  {Jaber Al-Edhari} A.,  {Ceccarelli} C.,  {L{\'o}pez-Sepulcre} A.,
   {Fontani} F.,   {Kama} M.,  2018, \mn@doi [\apj] {10.3847/1538-4357/aa9e88},
  \href {http://adsabs.harvard.edu/abs/2018ApJ...852..130K} {852, 130}

\bibitem[\protect\citeauthoryear{{Kaifu} et~al.,}{{Kaifu}
  et~al.}{2004}]{kaifu2004}
{Kaifu} N.,  et~al., 2004, \mn@doi [\pasj] {10.1093/pasj/56.1.69}, \href
  {http://adsabs.harvard.edu/abs/2004PASJ...56...69K} {56, 69}

\bibitem[\protect\citeauthoryear{{Keto}, {Caselli}  \& {Rawlings}}{{Keto}
  et~al.}{2015}]{keto2015}
{Keto} E.,  {Caselli} P.,   {Rawlings} J.,  2015, \mn@doi [\mnras]
  {10.1093/mnras/stu2247}, \href
  {http://adsabs.harvard.edu/abs/2015MNRAS.446.3731K} {446, 3731}

\bibitem[\protect\citeauthoryear{{Lis}, {Wootten}, {Gerin}  \& {Roueff}}{{Lis}
  et~al.}{2010}]{lis2010}
{Lis} D.~C.,  {Wootten} A.,  {Gerin} M.,   {Roueff} E.,  2010, \mn@doi [\apjl]
  {10.1088/2041-8205/710/1/L49}, \href
  {http://adsabs.harvard.edu/abs/2010ApJ...710L..49L} {710, L49}

\bibitem[\protect\citeauthoryear{{Magalhaes}, {Hily-Blant}, {Faure},
  {Hernandez-Vera}  \& {Lique}}{{Magalhaes} et~al.}{2018}]{magalhaes2018a}
{Magalhaes} V.,  {Hily-Blant} P.,  {Faure} A.,  {Hernandez-Vera} M.,   {Lique}
  F.,  2018, \aap\ in press

\bibitem[\protect\citeauthoryear{{Marty}, {Chaussidon}, {Wiens}, {Jurewicz}  \&
  {Burnett}}{{Marty} et~al.}{2011}]{marty2011}
{Marty} B.,  {Chaussidon} M.,  {Wiens} R.~C.,  {Jurewicz} A.~J.~G.,   {Burnett}
  D.~S.,  2011, \mn@doi [Science] {10.1126/science.1204656}, \href
  {http://adsabs.harvard.edu/abs/2011Sci...332.1533M} {332, 1533}

\bibitem[\protect\citeauthoryear{{Milam}, {Savage}, {Brewster}, {Ziurys}  \&
  {Wyckoff}}{{Milam} et~al.}{2005}]{milam2005}
{Milam} S.~N.,  {Savage} C.,  {Brewster} M.~A.,  {Ziurys} L.~M.,   {Wyckoff}
  S.,  2005, \mn@doi [\apj] {10.1086/497123}, 634, 1126

\bibitem[\protect\citeauthoryear{{Ohishi} \& {Kaifu}}{{Ohishi} \&
  {Kaifu}}{1998}]{ohishi1998}
{Ohishi} M.,  {Kaifu} N.,  1998, \mn@doi [Faraday Discussions]
  {10.1039/a801058g}, \href {http://adsabs.harvard.edu/abs/1998FaDi..109..205O}
  {109, 205}

\bibitem[\protect\citeauthoryear{{Osamura}, {Fukuzawa}, {Terzieva}  \&
  {Herbst}}{{Osamura} et~al.}{1999}]{osamura1999}
{Osamura} Y.,  {Fukuzawa} K.,  {Terzieva} R.,   {Herbst} E.,  1999, \mn@doi
  [\apj] {10.1086/307406}, \href
  {http://adsabs.harvard.edu/abs/1999ApJ...519..697O} {519, 697}

\bibitem[\protect\citeauthoryear{{Qu{\'e}nard}, {Vastel}, {Ceccarelli},
  {Hily-Blant}, {Lefloch}  \& {Bachiller}}{{Qu{\'e}nard}
  et~al.}{2017}]{quenard2017}
{Qu{\'e}nard} D.,  {Vastel} C.,  {Ceccarelli} C.,  {Hily-Blant} P.,  {Lefloch}
  B.,   {Bachiller} R.,  2017, \mn@doi [\mnras] {10.1093/mnras/stx1373}, \href
  {http://adsabs.harvard.edu/abs/2017MNRAS.470.3194Q} {470, 3194}

\bibitem[\protect\citeauthoryear{{Romano}, {Matteucci}, {Zhang}, {Papadopoulos}
   \& {Ivison}}{{Romano} et~al.}{2017}]{romano2017}
{Romano} D.,  {Matteucci} F.,  {Zhang} Z.-Y.,  {Papadopoulos} P.~P.,   {Ivison}
  R.~J.,  2017, \mn@doi [\mnras] {10.1093/mnras/stx1197}, \href
  {http://adsabs.harvard.edu/abs/2017MNRAS.470..401R} {470, 401}

\bibitem[\protect\citeauthoryear{{Roueff}, {Loison}  \& {Hickson}}{{Roueff}
  et~al.}{2015}]{roueff2015}
{Roueff} E.,  {Loison} J.~C.,   {Hickson} K.~M.,  2015, \mn@doi [\aap]
  {10.1051/0004-6361/201425113}, \href
  {http://adsabs.harvard.edu/abs/2015A%26A...576A..99R} {576, A99}

\bibitem[\protect\citeauthoryear{{Rubin} et~al.,}{{Rubin}
  et~al.}{2015}]{rubin2015a}
{Rubin} M.,  et~al., 2015, \mn@doi [Science] {10.1126/science.aaa6100}, \href
  {http://adsabs.harvard.edu/abs/2015Sci...348..232R} {348, 232}

\bibitem[\protect\citeauthoryear{{Sakai}, {Saruwatari}, {Sakai}, {Takano}  \&
  {Yamamoto}}{{Sakai} et~al.}{2010}]{sakai2010}
{Sakai} N.,  {Saruwatari} O.,  {Sakai} T.,  {Takano} S.,   {Yamamoto} S.,
  2010, \mn@doi [\aap] {10.1051/0004-6361/200913098}, \href
  {http://adsabs.harvard.edu/abs/2010A%26A...512A..31S} {512, A31}

\bibitem[\protect\citeauthoryear{{Schwarz} \& {Bergin}}{{Schwarz} \&
  {Bergin}}{2014}]{schwarz2014}
{Schwarz} K.~R.,  {Bergin} E.~A.,  2014, \mn@doi [\apj]
  {10.1088/0004-637X/797/2/113}, \href
  {http://adsabs.harvard.edu/abs/2014ApJ...797..113S} {797, 113}

\bibitem[\protect\citeauthoryear{{Shinnaka}, {Kawakita}, {Jehin}, {Decock},
  {Hutsem{\'e}kers}, {Manfroid}  \& {Arai}}{{Shinnaka}
  et~al.}{2016}]{shinnaka2016b}
{Shinnaka} Y.,  {Kawakita} H.,  {Jehin} E.,  {Decock} A.,  {Hutsem{\'e}kers}
  D.,  {Manfroid} J.,   {Arai} A.,  2016, \mn@doi [\mnras]
  {10.1093/mnras/stw2410}, \href
  {http://adsabs.harvard.edu/abs/2016MNRAS.462S.195S} {462, S195}

\bibitem[\protect\citeauthoryear{{Sims}, {Queffelec}, {Travers}, {Rowe},
  {Herbert}, {Karth{\"a}user}  \& {Smith}}{{Sims} et~al.}{1993}]{sims1993}
{Sims} I.~R.,  {Queffelec} J.-L.,  {Travers} D.,  {Rowe} B.~R.,  {Herbert}
  L.~B.,  {Karth{\"a}user} J.,   {Smith} I.~W.~M.,  1993, \mn@doi [Chemical
  Physics Letters] {10.1016/0009-2614(93)87091-G}, \href
  {http://adsabs.harvard.edu/abs/1993CPL...211..461S} {211, 461}

\bibitem[\protect\citeauthoryear{{Spezzano}, {Caselli}, {Bizzocchi}, {Giuliano}
   \& {Lattanzi}}{{Spezzano} et~al.}{2017}]{spezzano2017}
{Spezzano} S.,  {Caselli} P.,  {Bizzocchi} L.,  {Giuliano} B.~M.,   {Lattanzi}
  V.,  2017, \mn@doi [\aap] {10.1051/0004-6361/201731262}, \href
  {http://adsabs.harvard.edu/abs/2017A%26A...606A..82S} {606, A82}

\bibitem[\protect\citeauthoryear{{Takano} et~al.,}{{Takano}
  et~al.}{1998}]{takano1998}
{Takano} S.,  et~al., 1998, \aap, \href
  {http://adsabs.harvard.edu/abs/1998A%26A...329.1156T} {329, 1156}

\bibitem[\protect\citeauthoryear{{Taniguchi} \& {Saito}}{{Taniguchi} \&
  {Saito}}{2017}]{taniguchi2017c}
{Taniguchi} K.,  {Saito} M.,  2017, preprint, \href
  {http://adsabs.harvard.edu/abs/2017arXiv170608662T} {} (\mn@eprint {arXiv}
  {1706.08662})

\bibitem[\protect\citeauthoryear{{Taniguchi}, {Saito}  \& {Ozeki}}{{Taniguchi}
  et~al.}{2016}]{taniguchi2016}
{Taniguchi} K.,  {Saito} M.,   {Ozeki} H.,  2016, \mn@doi [\apj]
  {10.3847/0004-637X/830/2/106}, \href
  {http://adsabs.harvard.edu/abs/2016ApJ...830..106T} {830, 106}

\bibitem[\protect\citeauthoryear{{Taniguchi}, {Ozeki}  \& {Saito}}{{Taniguchi}
  et~al.}{2017}]{taniguchi2017b}
{Taniguchi} K.,  {Ozeki} H.,   {Saito} M.,  2017, \mn@doi [\apj]
  {10.3847/1538-4357/aa82ba}, \href
  {http://adsabs.harvard.edu/abs/2017ApJ...846...46T} {846, 46}

\bibitem[\protect\citeauthoryear{{Terzieva} \& {Herbst}}{{Terzieva} \&
  {Herbst}}{2000}]{terzieva2000}
{Terzieva} R.,  {Herbst} E.,  2000, \mn@doi [\mnras]
  {10.1046/j.1365-8711.2000.03618.x}, \href
  {http://adsabs.harvard.edu/abs/2000MNRAS.317..563T} {317, 563}

\bibitem[\protect\citeauthoryear{{Vastel}, {Phillips}  \& {Yoshida}}{{Vastel}
  et~al.}{2004}]{vastel2004}
{Vastel} C.,  {Phillips} T.~G.,   {Yoshida} H.,  2004, \mn@doi [\apjl]
  {10.1086/421265}, \href {http://adsabs.harvard.edu/abs/2004ApJ...606L.127V}
  {606, L127}

\bibitem[\protect\citeauthoryear{{Vastel}, {Ceccarelli}, {Lefloch}  \&
  {Bachiller}}{{Vastel} et~al.}{2014}]{vastel2014}
{Vastel} C.,  {Ceccarelli} C.,  {Lefloch} B.,   {Bachiller} R.,  2014, \mn@doi
  [\apjl] {10.1088/2041-8205/795/1/L2}, \href
  {http://adsabs.harvard.edu/abs/2014ApJ...795L...2V} {795, L2}

\bibitem[\protect\citeauthoryear{{Vastel}, {Kawaguchi}, {Qu{\'e}nard},
  {Ohishi}, {Lefloch}, {Bachiller}  \& {M{\"u}ller}}{{Vastel}
  et~al.}{2018}]{vastel2018}
{Vastel} C.,  {Kawaguchi} K.,  {Qu{\'e}nard} D.,  {Ohishi} M.,  {Lefloch} B.,
  {Bachiller} R.,   {M{\"u}ller} H.~S.~P.,  2018, \mn@doi [\mnras]
  {10.1093/mnrasl/slx197}, \href
  {http://adsabs.harvard.edu/abs/2018MNRAS.474L..76V} {474, L76}

\bibitem[\protect\citeauthoryear{{Vigren} et~al.,}{{Vigren}
  et~al.}{2012}]{vigren2012}
{Vigren} E.,  et~al., 2012, \mn@doi [\apj] {10.1088/0004-637X/757/1/34}, \href
  {http://adsabs.harvard.edu/abs/2012ApJ...757...34V} {757, 34}

\bibitem[\protect\citeauthoryear{{Wampfler}, {J{\o}rgensen}, {Bizzarro}  \&
  {Bisschop}}{{Wampfler} et~al.}{2014}]{wampfler2014}
{Wampfler} S.~F.,  {J{\o}rgensen} J.~K.,  {Bizzarro} M.,   {Bisschop} S.~E.,
  2014, \mn@doi [\aap] {10.1051/0004-6361/201423773}, \href
  {http://adsabs.harvard.edu/abs/2014A%26A...572A..24W} {572, A24}

\bibitem[\protect\citeauthoryear{{Watson}, {Anicich}  \& {Huntress}}{{Watson}
  et~al.}{1976}]{watson1976b}
{Watson} W.~D.,  {Anicich} V.~G.,   {Huntress} W.~T.,  1976, \apjl, 205, L165

\bibitem[\protect\citeauthoryear{{Wirstr{\"o}m} \& {Charnley}}{{Wirstr{\"o}m}
  \& {Charnley}}{2018}]{wirstrom2018}
{Wirstr{\"o}m} E.~S.,  {Charnley} S.~B.,  2018, \mn@doi [\mnras]
  {10.1093/mnras/stx3030}, \href
  {http://adsabs.harvard.edu/abs/2018MNRAS.474.3720W} {474, 3720}

\bibitem[\protect\citeauthoryear{{Wirstr{\"o}m}, {Charnley}, {Cordiner}  \&
  {Milam}}{{Wirstr{\"o}m} et~al.}{2012}]{wirstrom2012}
{Wirstr{\"o}m} E.~S.,  {Charnley} S.~B.,  {Cordiner} M.~A.,   {Milam} S.~N.,
  2012, \mn@doi [\apjl] {10.1088/2041-8205/757/1/L11}, \href
  {http://adsabs.harvard.edu/abs/2012ApJ...757L..11W} {757, L11}

\bibitem[\protect\citeauthoryear{{Zeng} et~al.,}{{Zeng}
  et~al.}{2017}]{zeng2017}
{Zeng} S.,  et~al., 2017, \mn@doi [\aap] {10.1051/0004-6361/201630210}, \href
  {http://adsabs.harvard.edu/abs/2017A%26A...603A..22Z} {603, A22}

\bibitem[\protect\citeauthoryear{{van Dishoeck}, {Bergin}, {Lis}  \&
  {Lunine}}{{van Dishoeck} et~al.}{2014}]{vandishoeck2014a}
{van Dishoeck} E.~F.,  {Bergin} E.~A.,  {Lis} D.~C.,   {Lunine} J.~I.,  2014,
  {Water: From Clouds to Planets}.
pp 835--858 (\mn@eprint {arXiv} {1401.8103}),
  \mn@doi{10.2458/azu_uapress_9780816531240-ch036}

\bibitem[\protect\citeauthoryear{{van der Tak}, {Black}, {Sch{\"o}ier},
  {Jansen}  \& {van Dishoeck}}{{van der Tak} et~al.}{2007}]{vandertak2007}
{van der Tak} F.~F.~S.,  {Black} J.~H.,  {Sch{\"o}ier} F.~L.,  {Jansen} D.~J.,
   {van Dishoeck} E.~F.,  2007, \mn@doi [\aap] {10.1051/0004-6361:20066820},
  \href {http://adsabs.harvard.edu/abs/2007A%26A...468..627V} {468, 627}

\makeatother
\end{thebibliography}

\appendix

\section{RADEX/MCMC calculations}

\begin{figure}
  \centering
  \includegraphics[width=\hsize]{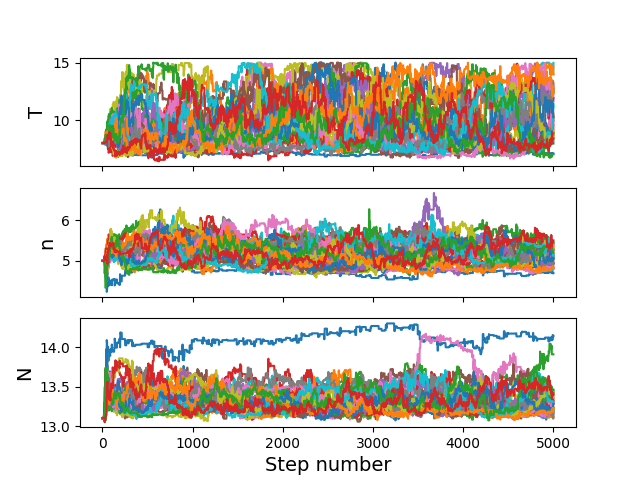}
  \includegraphics[width=\hsize]{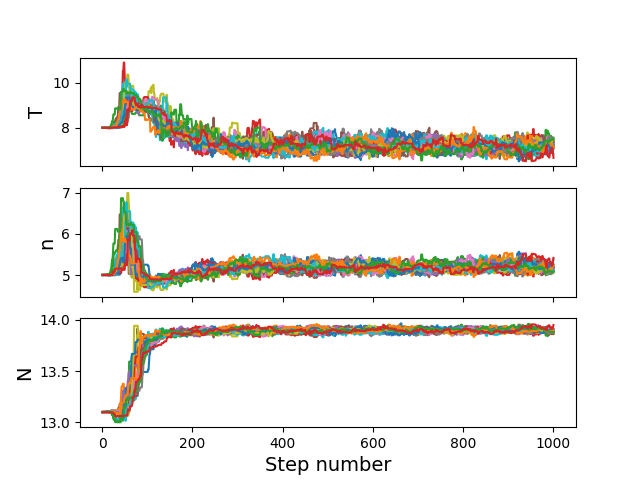}
  \caption{The 24 walkers along their 5000 and 1000 steps evolution
    associated with the rotational and hyperfine analysis (resp.) of
    the \ce{HC3N} lines (see Fig.~\ref{fig:mcmchc3n} and
    Table~\ref{tab:results}).}
  \label{fig:walkershc3nrot}
\end{figure}

\begin{figure}
  \centering
  \includegraphics[width=\hsize]{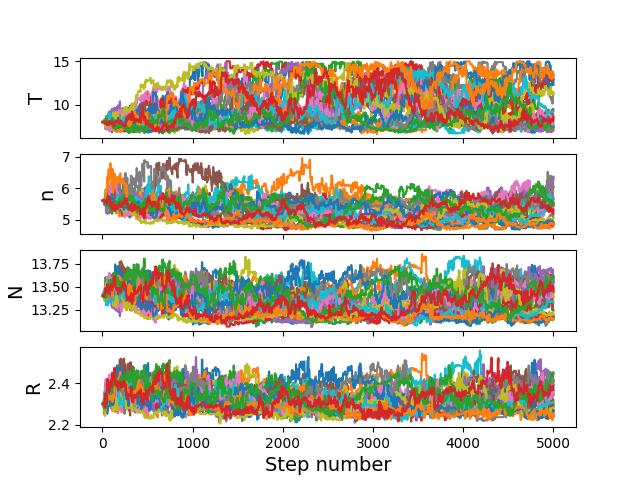}
  \includegraphics[width=\hsize]{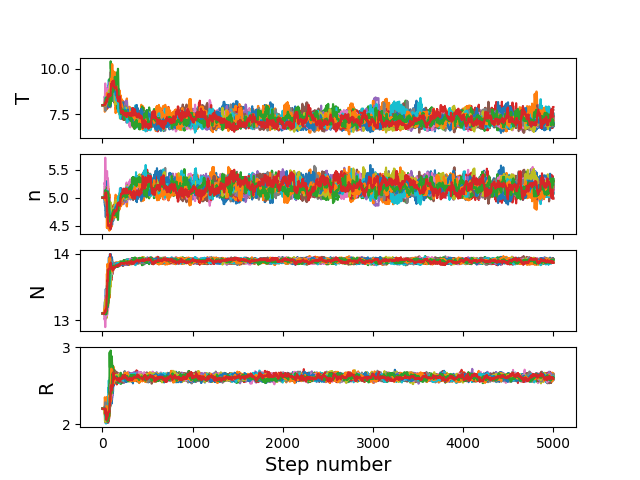}
  \caption{Markov chains of the runs used to measure the nitrogen
    isotopic ratio of \ce{HC3N} (see
    Table~\ref{tab:results}. \textit{Top:} all-rot
    results. \textit{Bottom:} all-hfs results. For the sake of
    clarity, the legends do not specify that the density and column
    density are on a log10 scale. Note that the prior probabilities
    are uniform for each parameter, within 6 to 15~K for the
    temperature, \dix{4} to \dix{12}\ccc\ for the \hh\ density, and
    \dix{12} to \dix{15}\cc\ for the column density of \ce{HC3N}.}
  \label{fig:mcmc15nwalkers}
\end{figure}

The walkers, or chains, in our rotational and hyperfine analysis of
the \ce{HC3N} lines are shown in \rfig{walkershc3nrot}. As can be
seen, a good convergence and parameter space exploration is obtained,
especially for the hyperfine analysis. Similar plots for the study of
the isotopic ratio are shown in \rfig{mcmc15nwalkers}. Note that the
prior probabilities are uniform for each parameter, within 6 to 15~K
for the kinetic temperature, \dix{4} to \dix{12}\ccc\ for the \hh\
density, and \dix{12} to \dix{15}\cc\ for the column density of
\ce{HC3N}. The hyperfine structure of \ce{HC3N} is detailed in
Table~\rtab{hc3nhfs}, where of the 6 hf transitions within each
rotational multiplet, only 5 are listed.

\begin{table}
  \centering
  \caption{\label{tab:hc3nhfs}Theoretical relative intensities of the
    various hf components of each rotational multiplet of \ce{HC3N}.}
  \begin{tabular}{rr crr}
    \toprule
    $N\ra N'^\dag$ & $F\ra F'^\dag$ & Frequency & $A_{ul}^\ddag$ & R.I.$^\S$ \\
    &&MHz & \pers & \% \\
    \hline
 8-7&   8-8&    72.78229&  4.60E-07&    0.52\\
 8-7&   7-6&    72.78381&  2.89E-05&   28.89\\
 8-7&   8-7&    72.78382&  2.90E-05&   32.85\\
 8-7&   9-8&    72.78383&  2.94E-05&   37.22\\
 8-7&   7-8&    72.78401&  2.04E-09&    0.00\\
 8-7&   7-7&    72.78554&  5.21E-07&    0.52\\
    \hline
 9-8&   9-9&    81.87992&  5.20E-07&    0.41\\
 9-8&   8-7&    81.88145&  4.16E-05&   29.42\\
 9-8&   9-8&    81.88147&  4.16E-05&   32.88\\
 9-8&  10-9&    81.88148&  4.22E-05&   36.87\\
 9-8&   8-9&    81.88163&  1.80E-09&    0.00\\
 9-8&   8-8&    81.88317&  5.82E-07&    0.41\\
    \hline
10-9&  10-10&    90.97744&  5.81E-07&    0.33\\
10-9&   9-8&    90.97898&  5.75E-05&   29.84\\
10-9&  10-9&    90.97899&  5.75E-05&   32.99\\
10-9&  11-10&    90.97900&  5.81E-05&   36.50\\
10-9&   9-10&    90.97914&  1.61E-09&    0.00\\
10-9&   9-9&    90.98069&  6.42E-07&    0.33\\
    \hline
11-10&  11-11&   100.07483&  6.42E-07&    0.28\\
11-10&  10-9&   100.07638&  7.70E-05&   30.15\\
11-10&  11-10&   100.07639&  7.71E-05&   33.07\\
11-10&  12-11&   100.07640&  7.77E-05&   36.22\\
11-10&  10-11&   100.07652&  1.46E-09&    0.00\\
11-10&  10-10&   100.07808&  7.03E-07&    0.28\\
    \hline
12-11&  12-12&   109.17208&  7.03E-07&    0.23\\
12-11&  12-11&   109.17364&  1.01E-04&   33.28\\
12-11&  11-10&   109.17364&  1.00E-04&   30.31\\
12-11&  13-12&   109.17365&  1.01E-04&   35.94\\
12-11&  11-11&   109.17532&  7.64E-07&    0.23\\
    \bottomrule
  \end{tabular}
  \begin{list}{}{}
  \item[$\dag$] $N$ is the rotational quantum number, $F=N, N\pm I$
    identifies the hyperfine sub-levels, and $I=1$ is the nuclear spin
    of the \foun\ atom.
  \item[$\ddag$] Hyperfine Einstein coefficient for spontaneous decay.
  \item[$\S$] Normalized relative intensities in percent.
  \end{list}
\end{table}
\subsection{Physical conditions}

The predicted flux and flux ratios corresponding to our best models
are summarized in \rtab{check} and shown in \rfig{results}.

\begin{table*}
  \centering
  \caption{\label{tab:check}Model predictions for the most probable
    solutions from the rotational and hyperfine (fluxes and flux
    ratios) analysis.}
  \begin{tabular}{lc cc cc cc}
    \hline
    Species & Transition$^\S$ & \mc{2}{Observations$^\dag$}
    &\mc{2}{all-rot$^\ddag$}
    &\mc{2}{all-hfs$^\ddag$}\\
    && $W$ & $\sigma$
    &$\widetilde{W}$ & $|\widetilde{W}-W|/\sigma$
    &$\widetilde{W}$ & $|\widetilde{W}-W|/\sigma$\\
    \hline
    \ce{HC3N}
    & 8-7
      &  37.9 &        1.9 &       17.8 &      -10.6 &       46.4 &        4.5\\
    & &2053.0 &      102.7 &     1971.3 &       -0.8 &     1982.4 &       -0.7\\
    & &  50.3 &        2.7 &       17.8 &      -12.1 &       46.9 &       -1.3\\
    & 9-8
      &  15.9 &        2.2 &       10.9 &       -2.3 &       25.2 &        4.2\\
    & &1633.0 &       81.7 &     1662.4 &        0.4 &     1756.9 &        1.5\\
    & &  21.2 &        2.2 &       10.9 &       -4.7 &       25.5 &        2.0\\
    & 10-9
      &  16.4 &        0.8 &        6.2 &      -12.5 &       12.3 &       -5.0\\
    & &1404.0 &       70.2 &     1295.9 &       -1.5 &     1474.3 &        1.0\\
    & &  16.3 &        0.8 &        6.2 &      -12.4 &       12.4 &       -4.7\\
    & 11-10
      &   6.4 &        0.7 &        3.2 &       -4.5 &        5.4 &       -1.4\\
    & & 956.2 &       47.8 &      915.6 &       -0.8 &     1091.6 &        2.8\\
    & &  10.3 &        0.8 &        3.2 &       -9.0 &        5.5 &       -6.2\\
    & 12-11
      & 626.1 &       31.3 &      571.9 &       -1.7 &      670.7 &        1.4\\
    \ce{HC3^{15}N}
    & 9-8  &  14.1 &        2.1 &       12.4 &       -0.8 &       14.8 &        0.4\\
    &10-9  &   9.0 &        0.7 &        8.8 &       -0.3 &        9.1 &        0.1\\
    &11-10 & $<$7  &        1.4 &        5.8 &        3.7 &        5.0 &        3.2\\
    &12-11 & $<$5  &        1.0 &        3.4 &        1.0 &        2.5 &        0.0\\
    &13-12 & $<$22 &        4.3 &        1.8 &       -0.8 &        1.1 &       -1.0\\
    \hline
  \end{tabular}
  \begin{list}{}{}
  \item $^\S$ Hyperfine transitions of \ce{HC3N} are those of
    Table~\ref{tab:hc3nhfs}, starting with the main hf group, followed
    by the other two hf lines, sorted by increasing frequency.
  \item $^\dag$ The adopted uncertainties are at least 5\% of the
    flux. Upper limits are at 5$\sigma$. All integrated intensities
    are in mK \kms.
  \item $^\ddag$ $\widetilde{W}$ is the predicted flux {based
      on the models given in Table~\ref{tab:results}.}
  \end{list}
\end{table*}

\bsp	
\label{lastpage}
\end{document}